\pdfoutput=1
\documentclass[prodmode,acmtweb]{arxivacmsmall} 

\usepackage{hyperref}

\usepackage{subfig}
\usepackage{multirow}

\usepackage[ruled]{algorithm2e}
\usepackage{algorithmic}

\usepackage[colorinlistoftodos]{todonotes}

\usepackage{siunitx}
\usepackage{booktabs}
\newcommand{\microop}{$\mu$op}
\newcommand{\microops}{\microop{}s}
\usepackage{microtype}
\usepackage{tikz}

\usetikzlibrary{chains,arrows,automata,decorations.markings,positioning,calc,decorations.pathreplacing,patterns}

\usepackage{color}
    \definecolor{lightgray}{rgb}{0.95, 0.95, 0.95}
    \definecolor{darkgray}{rgb}{0.4, 0.4, 0.4}
    \definecolor{purple}{rgb}{0.65, 0.12, 0.82}
    \definecolor{ocherCode}{rgb}{1, 0.5, 0} 
    \definecolor{blueCode}{rgb}{0, 0, 0.93} 
    \definecolor{greenCode}{rgb}{0, 0.6, 0} 
\usepackage{csquotes}
\usepackage{listings}
\lstdefinestyle{customc}{%
  belowcaptionskip=1\baselineskip,
  breaklines=true,
  xleftmargin=\parindent,
  language=C,
  showstringspaces=false,
  basicstyle=\small\ttfamily,
  keywordstyle=\bfseries\color{green!40!black},
  numberstyle=\tiny,
  commentstyle=\itshape\color{purple!40!black},
  identifierstyle=\bfseries\color{black},
  stringstyle=\color{orange},
   morekeywords={uint64_t,uint32_t,__m256i,__m128i,UINT64_C},
}

\lstdefinestyle{htmlCode} {
   language=html,
   basicstyle=\scriptsize\ttfamily,
   keywordstyle=\bfseries\ttfamily,
   commentstyle=\color{gray}\ttfamily,
    identifierstyle=\color{black},
    keywordstyle=\color{black}\bfseries,
    ndkeywordstyle=\color{greenCode}\bfseries,
    stringstyle=\color{black}\bfseries,
    commentstyle=\color{darkgray}\ttfamily,
   escapechar=| 
}
\usepackage[english]{babel}
\usepackage[utf8x]{inputenc}
\usepackage{amsmath}
\usepackage{url}

\SetAlFnt{\small}
\SetAlCapFnt{\small}
\SetAlCapNameFnt{\small}
\SetAlCapHSkip{0pt}
\IncMargin{-\parindent}
\newcommand{\opcode}[1]{\texttt{#1}}
\newcommand{\intrinsic}[1]{\texttt{#1}}

\usepackage{etoolbox}

\makeatletter
\patchcmd{\maketitle}{\@bottomstuff}{}{}{}

\patchcmd{\maketitle}{\@copyrightspace}{}{}{}
\def\runningfoot{\def\@runningfoot{}}
\def\firstfoot{\def\@firstfoot{}}
\makeatother
\usepackage{fancyhdr}
\begin{document}

\markboth{W. Mu\l a and D. Lemire}{Faster Base64 Encoding and Decoding using AVX2 Instructions}

\title{Faster Base64 Encoding and Decoding using AVX2 Instructions}

\author{WOJCIECH MU\L A
\affil{~}
DANIEL LEMIRE
\affil{Universit\'e du Qu\'ebec (TELUQ)}
}

\begin{abstract}
Web developers use base64 formats to include images, fonts, sounds and other resources directly inside HTML, JavaScript, JSON and XML files. We estimate that billions of base64 messages are decoded every day. We are motivated to improve the efficiency of base64 encoding and decoding. Compared to state-of-the-art implementations, we multiply  the speeds of both the  encoding ($\approx 10 \times$) and the decoding ($\approx 7 \times$). We achieve these good results by  using the single-instruction-multiple-data (SIMD) instructions available on recent Intel processors (AVX2). Our accelerated software abides by the specification and reports errors when encountering characters outside of the base64 set. It is available online as free software under a liberal license.
\end{abstract}

\ccsdesc[500]{Theory of computation~Vector / streaming algorithms}
%
%

\terms{Algorithms, Performance}

\keywords{Binary-to-text encoding, Vectorization, Data URI, Web Performance}

\acmformat{Wojciech Mu\l a,  Daniel Lemire, 2017. Faster Base64 Encoding and Decoding using AVX2 Instructions.}

\begin{bottomstuff}
This work is supported by Natural Sciences and Engineering Research Council of Canada,  grant 261437.

Author's addresses: D. Lemire, Universit\'e du Qu\'ebec (TELUQ), 5800,  Saint-Denis street, Montreal (Quebec)  H2S 3L5, Canada.
\end{bottomstuff}

\maketitle

\section{Introduction}

We use base64 formats to represent arbitrary binary data as text. Base64 is part of the MIME email protocol~\cite{rfc1421,rfc2045}, used to encode binary attachments. Base64 is included in the standard libraries of popular programming languages such as Java, C\#, Swift, PHP, Python,  Rust, JavaScript and Go. 
Major database systems such as Oracle and MySQL include base64 functions.

On the Web, we often combine binary resources (images, videos, sounds) with text-only documents (XML, JavaScript, HTML). Before a Web page can be displayed, it is often necessary to retrieve not only the HTML document but also all of the separate binary resources it needs. The round-trips needed to retrieve all of the resources are often a performance bottleneck~\cite{Everts:2013:RMP:2492007.2492024}. Consequently, major websites---such as Google, Bing, and Baidu---deliver  small images within HTML pages using the data URI scheme~\cite{rfc2397}. A data URI takes the form ``\texttt{data:<content type>:;base64,<base64 data>}''. For example, consider the \texttt{img} element
\begin{lstlisting}[style=htmlCode] 
<img 
src="data:image/gif;base64,R0lGODlhAQABAIAAAP///wAAACwAAAAAAQABAAACAkQBADs=" />
\end{lstlisting}
where the text ``\texttt{R0lGODl}\ldots'' is a base64 representation of the binary data of a GIF image.
 Data URIs are supported by all major browsers~\cite{Johansen:2013:CBS:2487788.2487797}. We estimate that billions of pages containing base64 data are loaded every day. 
 
Base64 formats encode arbitrary bytes into a stream of characters chosen from a list of 64~ASCII characters. Three arbitrary bytes can be thus encoded using four ASCII characters. Though base64 encoding increases the number
of bytes by 33\%, this is alleviated by the commonly used text compression included in the HTTP protocol~\cite{rfc2616}. The size difference, after compression, can be much smaller than 33\% and might even be negligible~\cite{Calhoun2011}.

Base64 has many applications on the Web beyond embedding resources within HTML pages as an optimization:
\begin{itemize}
\item 
 The recently introduced Web Storage 
  specification allows Web developers to store text data (including base64-encoded resources) persistently within the browser~\cite{Hickson2016}. With Web Storage, developers can ensure that base64-encoded images and fonts are cached in the browser.
\item 
Similarly, base64  embeds binary data within XML and JSON files generated by web services, as these text-only formats do not otherwise allow binary content. A Web page can retrieve XML and JSON documents and decode the corresponding dynamically-generated binary resources on the fly.
Correspondingly, several database systems frequently code and decode  base64 strings even though they store binary data as binary:
\begin{itemize}
\item MongoDB normally receives and sends binary data as base64-encoded strings~\cite{MongoDBJSON}.
\item Elasticsearch accepts binary values as  base64-encoded strings~\cite{ElasticsearchJSON}. 
\item SQL Server users can add the \texttt{BINARY BASE64} qualifier when issuing \texttt{FOR XML} queries, so that the generated XML encodes binary objects using base64~\cite{sqlserverbase64}. 
\item Amazon SimpleDB automatically encodes data sequences that are not valid in XML using base64~\cite{simpledb}.
\item Amazon DynamoDB supports binary attributes, but they are normally exchanged in a base64-encoded form within JSON documents~\cite{DynamoDB}.
Crane and Lin report that decoding binary attributes from base64 is slow~\cite{Crane:2017:ESA:3121050.3121086}.
\end{itemize}

\end{itemize}

\noindent Base64 can also be used for security and privacy purposes. Credentials are often stored and transmitted using base64, e.g., in the HTTP Basic authentication method. There are also more advanced applications: 
\begin{itemize}
\item 
Many systems allow users to communicate text more freely than binary data.
Using this principle, Tierney et al.\ use base64 to allow users to share encrypted pictures on social networks~\cite{Tierney:2013:CPP:2512938.2512939}, even when such networks do not natively support this feature.
\item  Moreover, even when multiple HTTP queries to retrieve resources are efficient, they make it easier for adversaries to track users. Indeed, TCP/IP packet headers cannot be encrypted and they reveal the size of the data, as well as the destination and source addresses. Thus even encrypted Web access may not guarantee anonymity.
Tang and Lin show that we can use base64 to better obfuscate Web queries~\cite{tang2015eqpo}.
\end{itemize}

Encoding and decoding base64 data is fast. We do not expect base64 decoding to be commonly a bottleneck in Web browsers. Yet it can still be much slower to decode data than to copy it: e.g., \texttt{memcpy} may use as little as 0.03~cycles per byte  while a fast base64 decoder might use  1.8~cycles per byte on the same test (and be $60\times$ slower), see Table~\ref{tab:perf}.  Because base64 is ubiquitous and used on a massive scale within servers and database systems, there is industry interest in  making it run faster~\cite{aws2014}. 

Most commodity processors (Intel, AMD, ARM, POWER) benefit from 
single-instruction-multiple-data (SIMD) instructions. Unlike regular (scalar) instructions, these SIMD instructions operate on several words at once (or ``vectors'').  Though compilers can automatically use these instructions, it may be necessary to design algorithms with SIMD instructions in mind for best speed.
Unlike regular (or ``scalar'') instructions operating on single words, SIMD instructions operate
  on several words at once. We refer to these groups of words as vectors. These vectors are implemented as wide registers within the processors. For example, recent x64 processors benefit from AVX2
  instructions, operating on 256-bit vectors. We treat such vectors  as arrays of 32~bytes, arrays of sixteen 16-bit integers or arrays of eight 32-bit integers.

\section{Base64}
\label{sec:base64}

Base64 code is made streams of 6-bit words represented as ASCII characters. Blocks of four 6-bit words correspond bijectively to blocks of three 8-bit words (bytes).  
\begin{itemize}
\item During the encoding
of an arbitrary binary stream, each block of three input bytes (or $3\times 8=24$~bits) is unpacked to four~6-bit words ($4\times 6=24$~bits).  Each of the four 6-bit words corresponds to an ASCII character. See Algorithm~\ref{algo:base64encoding}. 
If the
length of the input is not divisible by three bytes, then the encoder may use the special 
padding character ('\texttt{=}'). There is one padding character per leftover byte (one or two). The length of a valid
base64 string is normally divisible by four.  In some applications, it may be acceptable to omit the padding characters ('\texttt{=}') if the size of the binary data is otherwise known. 
\item 
Most base64 decoders translate blocks of four ASCII letters into blocks of four 6-bit integer values (in $[0, 63)$). Each of these blocks is then packed into three bytes. See Algorithm~\ref{algo:base64decoding}. When the base64 stream ends with one or two padding characters ('\texttt{=}'), two or one final bytes are decoded.
\end{itemize}

\begin{algorithm}
\begin{algorithmic}[1]
\REQUIRE  A stream $s$ of $n$ bytes, indexed as $s_0, s_1, \ldots, s_{n-1} \in [0,256)$
\REQUIRE  A function $B$ mapping values in $[0,64)$ to ASCII characters (e.g., see Table~\ref{tab:base64table})
\STATE $p\leftarrow$ empty buffer of ASCII characters
\FOR{$i$ in $0,3, \ldots, n - (n \mod 3) - 3$ }
\STATE append $B(s_i \div 4)$ to $p$ 
\STATE append $B(((s_i \times 16) \bmod 64) + (s_{i+1} \div 16))$ to $p$ 
\STATE append $B(((s_{i+1} \times 4) \bmod 64) + (s_{i+2} \div 64))$ to $p$ 
\STATE append $B((s_{i+2} \bmod 64) )$ to $p$ 
\ENDFOR
\STATE $i \leftarrow n - n \mod 3$
\IF{$i < n$}
\STATE append $B(s_i \div 4)$ to $p$ 
\IF{$i = n - 1$}
\STATE append $B(((s_i \times 16) \bmod 64) )$ to $p$ 
\STATE append padding character '\texttt{=}' to $p$
\ELSIF {$i = n - 2$}
\STATE append $B(((s_i \times 16) \bmod 64) + (s_{i+1} \div 16))$ to $p$ 
\STATE append $B(((s_{i+1} \times 4) \bmod 64))$ to $p$ 
\ENDIF 
\STATE append padding character '\texttt{=}' to $p$
\ENDIF 
\STATE \textbf{return} $p$
\end{algorithmic}
\caption{Base64 encoding \label{algo:base64encoding}}
\end{algorithm}

\begin{algorithm}
\begin{algorithmic}[1]
\REQUIRE  A stream $c$ of $n$ ASCII characters, indexed as $C_0, C_1, \ldots, c_{n-1}$, $n$ must be divisible by 4
\REQUIRE  A function $A$ mapping ASCII characters to values in  $[0,64)$ (e.g., see Table~\ref{tab:base64table}), using the conventional that the padding character '\texttt{=}' has value 0, and returning a negative integer if an unsupported ASCII character is found 
\STATE $p\leftarrow$ empty buffer of bytes used to store values in $[0,256)$
\FOR{$i$ in $0,4, \ldots, n - 4$ }
\STATE $a \leftarrow A(C_i)$
\STATE $b \leftarrow A(C_{i+1})$
\STATE $c \leftarrow A(C_{i+2})$
\STATE $d \leftarrow A(C_{i+3})$
\IF {any of $a,b,c,d$ is negative}
\STATE report an error as  unexpected character was encountered (based on Table~\ref{tab:base64table})
\ENDIF
\STATE append byte value $(a \times 4) + (b \div 16)$ to $p$ 
\IF {$C_{i+2}$ is the padding character ('\texttt{=}') }
\STATE \textbf{return} $p$
\ENDIF
\STATE append byte value $(b \times 16) \bmod 256 + (c \div 4)$ to $p$ 
\IF {$C_{i+3}$ is the padding character ('\texttt{=}') }
\STATE \textbf{return} $p$
\ENDIF
\STATE append byte value $(c \times 64) \bmod 256 + d$ to $p$ 
\ENDFOR
\STATE \textbf{return} $p$
\end{algorithmic}
\caption{Base64 decoding \label{algo:base64decoding}}
\end{algorithm}

Base64 standards define a lookup table to translate between
6-bit values (in $[0,63)$) and ASCII characters. We consider the standard~\cite{rfc4648}
where the following characters are used: \texttt{A} \ldots \texttt{Z}, \texttt{a} \ldots \texttt{z},
\texttt{0} \ldots \texttt{9}, \texttt{+} and \texttt{/}, as  in Table~\ref{tab:base64table}. 
Unless otherwise specified, the decoder should  report an error when characters outside of this set are encountered.

\begin{table}
\tbl{Base64 mapping\label{tab:base64table} between 6-bit values and ASCII characters. For each ASCII character, we also provide the code point or byte value as an hexadecimal number. The '\texttt{=}' character  pads the end of the stream if the number of  bytes is not divisible by 3.}{%
\begin{tabular}{ccc|ccc|ccc|ccc}
\toprule
value & ASCII & char & value & ASCII & char & value & ASCII & char & value & ASCII & char \\ 
\midrule
0 & 0x41 & \texttt{A} & 16 & 0x51 & \texttt{Q} & 32 & 0x67 & \texttt{g} & 48 & 0x77 & \texttt{w} \\
1 & 0x42 & \texttt{B} & 17 & 0x52 & \texttt{R} & 33 & 0x68 & \texttt{h} & 49 & 0x78 & \texttt{x} \\
2 & 0x43 & \texttt{C} & 18 & 0x53 & \texttt{S} & 34 & 0x69 & \texttt{i} & 50 & 0x79 & \texttt{y} \\
3 & 0x44 & \texttt{D} & 19 & 0x54 & \texttt{T} & 35 & 0x6a & \texttt{j} & 51 & 0x7a & \texttt{z} \\
4 & 0x45 & \texttt{E} & 20 & 0x55 & \texttt{U} & 36 & 0x6b & \texttt{k} & 52 & 0x30 & \texttt{0} \\
5 & 0x46 & \texttt{F} & 21 & 0x56 & \texttt{V} & 37 & 0x6c & \texttt{l} & 53 & 0x31 & \texttt{1} \\
6 & 0x47 & \texttt{G} & 22 & 0x57 & \texttt{W} & 38 & 0x6d & \texttt{m} & 54 & 0x32 & \texttt{2} \\
7 & 0x48 & \texttt{H} & 23 & 0x58 & \texttt{X} & 39 & 0x6e & \texttt{n} & 55 & 0x33 & \texttt{3} \\
8 & 0x49 & \texttt{I} & 24 & 0x59 & \texttt{Y} & 40 & 0x6f & \texttt{o} & 56 & 0x34 & \texttt{4} \\
9 & 0x4a & \texttt{J} & 25 & 0x5a & \texttt{Z} & 41 & 0x70 & \texttt{p} & 57 & 0x35 & \texttt{5} \\
10 & 0x4b & \texttt{K} & 26 & 0x61 & \texttt{a} & 42 & 0x71 & \texttt{q} & 58 & 0x36 & \texttt{6} \\
11 & 0x4c & \texttt{L} & 27 & 0x62 & \texttt{b} & 43 & 0x72 & \texttt{r} & 59 & 0x37 & \texttt{7} \\
12 & 0x4d & \texttt{M} & 28 & 0x63 & \texttt{c} & 44 & 0x73 & \texttt{s} & 60 & 0x38 & \texttt{8} \\
13 & 0x4e & \texttt{N} & 29 & 0x64 & \texttt{d} & 45 & 0x74 & \texttt{t} & 61 & 0x39 & \texttt{9} \\
14 & 0x4f & \texttt{O} & 30 & 0x65 & \texttt{e} & 46 & 0x75 & \texttt{u} & 62 & 0x2b & \texttt{+} \\
15 & 0x50 & \texttt{P} & 31 & 0x66 & \texttt{f} & 47 & 0x76 & \texttt{v} & 63 & 0x2f & \texttt{/} \\

\bottomrule
\end{tabular}}
\end{table}

Sometimes, we want to encode binary data within an URL where the '\texttt{+}' and '\texttt{/}' characters have special meaning. Thus we may choose an instance of base64 called base64url~\cite{rfc4648}. The sole difference is that value 62 is represented by '\texttt{-}'  instead of '\texttt{+}' and the value 63 is represented by '\texttt{\_}' instead of '\texttt{/}'. Thus base64url avoids using the characters '\texttt{+}' and '\texttt{/}', and a base64url text can be safely included in an URL\@.  The JSON Web Signature proposal relies on base64url~\cite{rfc7515}. Our work would be equally applicable to base64url, as the difference between base64 and base64url has little impact on encoding and decoding algorithms.

\subsection{Character Encodings}

Base64 was designed with the ASCII character encoding in mind~\cite{rfc4648}. In a document using the ASCII encoding, only seven of the eight bits of each byte is used. By convention, each ASCII character has a corresponding byte value (also called code point) in $[0,128)$. 

  There are several supersets to the ASCII character encoding (e.g., UTF-8 or ISO~8859-1): they interpret strings of byte values in $[0,128)$ as ASCII strings. Only byte values with the most significant bits set are interpreted differently (e.g., as accented characters such as '\'e'). In other words, if we need to include an ASCII string within a string that uses a superset of the ASCII character encoding, we only need to copy the byte values. Thus base64 is practical with all ASCII supersets.
 
Most  Web pages are served using the Unicode format UTF-8~\cite{Google2012} which supports up to \num{1114112}~possible characters. Some programming languages (e.g., Go and Python) 
also default on UTF-8. XML documents use UTF-8 by default.
 Conveniently, UTF-8 is an ASCII superset. In UTF-8, only the ASCII characters can be represented using a single byte. All non-ASCII characters in UTF-8 require from two to four bytes.
 
 It might seem like base64 is suboptimal: there are many more than 64~distinct characters.  However, there are only 95 printable ASCII characters, and they include the space and the quotes (\texttt{"} and \texttt{'}), the ampersand (\texttt{\&}) and the less-than sign (\texttt{<}). Thus there are only about 90~characters that are represented as a single byte in UTF-8 that would be generally usable in HTML and XML\@. If we restrict the size of our table to a power of two, for simplicity and computational efficiency, then 64~characters is best.

\subsection{Efficient Scalar Encoding}

Throughout, we consider byte values as unsigned integers in $[0, 256)$. Thus we can think of  encoding as mapping a stream of numbers in $[0, 256)$ to ASCII characters.

In the main loop of the encoding Algorithm~\ref{algo:base64encoding}, we combine three-byte values ($s_i, s_{i+1}, s_{i+2}$) arithmetically into four values in $[0,64)$, that is
\begin{itemize}
\item  $s_i \div 4$,
\item $((s_i \times 16) \bmod 64) + (s_{i+1} \div 16)$,
\item   $((s_{i+1} \times 4) \bmod 64) + (s_{i+2} \div 64)$ 
\item and 
$s_{i+2} \bmod 64 $.
\end{itemize}
Then we pass these four values to the function $B$ which looks up the corresponding ASCII character.
In practice, an encoder might implement the function $B$ efficiently  as a lookup table, using a 64-byte array.

Thus, given three input bytes, we get four output characters:
\begin{itemize}
\item The first character is generated from the six most significant bits of the first byte value.
\item  The second character is determined from the two least significant bits from the first byte value and from  the four most significant bits of the second byte value.
\item The third character represents the four least significant bits of the second byte value and the two most significant bits from the third byte value.
\item The last character is determined by the six least significant bits from the third byte value.
\end{itemize}
Thus if we write the byte values starting from the left, with the bits within each byte written from the most significant to the least significant, we have that the first 6~bits from the left determine the first character, the next 6~bits the second character and so forth.

\begin{example}
Suppose that we need to encode the three byte values 71, 73, 70. We construct four 6-bit values out of these four bytes:
\begin{itemize}
\item  $71 \div 4 = 17$,
\item $((71 \times 16) \bmod 64) + (73 \div 16)= 52$,
\item   $((73 \times 4) \bmod 64) + (70 \div 64) = 37$ 
\item and 
$70 \bmod 64 = 6$.
\end{itemize}
We then look-up the resulting 6-bit values 17, 52, 37, 6 in Table~\ref{tab:base64table}
to get the ASCII characters \texttt{R}, \texttt{0}, \texttt{l}, \texttt{G}. The string  \texttt{R0lG} is the base64-encoded version of the three input bytes 71, 73, 70.
\end{example}

One of the fastest encoder~\cite{Galbreathfast} (used by the Google Chrome browser) optimizes the main loop of Algorithm~\ref{algo:base64encoding} by using several  256-byte arrays. One 256-byte array represents
$x\to B(x \div 4)$, whereas another 
represents $x\to B(x \bmod 64)$. This saves a few operations at the expense of a slightly increased memory usage:
\begin{itemize}
\item  instead of computing $s_i \div 4$ and then looking up the resulting index in a table, one can seek directly $s_i$ in a larger table, saving the cost of the division by 4 (which can be implemented as a shift);
\item instead of computing $s_{i+2} \bmod 64$ before looking up the result in a table, we can seek $s_{i+2}$ directly in a larger table, saving the cost of the modulo reduction (which can be implemented with a bitwise AND).
\end{itemize}
Using additional memory to save a few arithmetic and logical operations may improve the performance.

\subsection{Efficient Scalar Decoding}

We assume that the input base64 data is encoded in ASCII or  some ASCII superset such as  UTF-8 or ISO~8859-1. In such cases, any non-ASCII character should trigger an error. If there are white-space characters (e.g., '\verb+ +', '\verb+\n+' or  '\verb+\r+'), then they must be removed prior to decoding (see  Appendix~\ref{appendix:white}).

Algorithm~\ref{algo:base64decoding} illustrates a decoding procedure. The main loop of the algorithm consists of two steps, we first map ASCII characters to 6-bit integer values:
\begin{itemize}
\item  $a \leftarrow A(C_i)$,
\item  $b \leftarrow A(C_{i+1})$,
\item  $c \leftarrow A(C_{i+2})$,
\item  $d \leftarrow A(C_{i+3})$.
\end{itemize}
And then we compute the three output byte values:
\begin{itemize}
\item  $(a \times 4) + (b \div 16)$,
\item  $(b \times 16) \bmod 256 + (c \div 4)$,
\item  $(c \times 64) \bmod 256 + d$.
\end{itemize}
The function $A$ can be implemented as a table lookup.

\begin{example}
Let us consider the following base64 code \texttt{R0lGODlhAQABAIAAAP///wAAACwAAAAAAQABAAACAkQBADs=}. It represents a small gif image. To decode it, we may look up each character for the base64 code (except the terminating '\texttt{=}') in Table~\ref{tab:base64table} and map it to its corresponding value in $[0,64)$: 17, 52, 37, 6, 14, 3, 37, 33, 0, 16, 0, 1, 0, 8, 0, 0, 0, 15, 63, 63, 63, 48, 0, 0, 0, 2, 48, 0, 0, 0, 0, 0, 0, 16, 0, 1, 0, 0, 0, 2, 0, 36, 16, 1, 0, 3, 44. Then we take each block of 4~consecutive values $a, b, c, d$ and map them to $(a \time 4) + (b \div 16), (b \times 16) \bmod 256 + (c \div 4), (c \times 64) \bmod 256 + d$. We must pay attention to the fact that we are missing one value in the last block because of the terminating  '\texttt{=}' which means that, in the processing of this last block, we must produce only two values, instead of three.
The  result is the byte values:
71, 73, 70, 56, 57, 97, 1, 0, 1, 0, 128, 0, 0, 255, 255, 255, 0, 0, 0, 44, 0, 0, 0, 0, 1, 0, 1, 0, 0, 2, 2, 68, 1, 0, 59.
\end{example}

The fast decoder used by Google Chrome browser has a streamlined approach for all but the four last input characters. Instead of a single function $A$, it uses four distinct lookup tables ($A_1, A_2, A_3, A_4$) made of 256 32-bit values:
 $a \leftarrow A_1(C_i)$,
 $b \leftarrow A_2(C_{i+1})$,
$c \leftarrow A_3(C_{i+2})$,
 $d \leftarrow A_4(C_{i+3})$.
 Normally, only three of four bytes of each of these 32-bit values are used, with the remaining byte set to zero. However, whenever an illegal character is encountered, the extra byte is used as a flag.  We compute the bitwise OR of the four 32-bit values ($a,b,c,d$): $z = a \lor b \lor c \lor d$. We choose $A_1, A_2, A_3, A_4$ so that  three bytes of the 32-bit value $z$ are the decoded bytes. In effect, this approach decodes 4~input ASCII characters using four lookup and three bitwise OR, not counting a test for  illegal characters.
 The details depend on whether the hardware has a big endian or little endian architecture. This approach uses more memory, but it results in fewer operations and potentially higher speed. 

\section{Advanced x64 Instructions and Intrinsics}
\label{sec:advancedinstructions}

Commodity x64 processors benefit from several advanced instructions. Recent compilers can automatically make use of them without any intervention from  the programmer.

However, our experience is that programmers who tune their code to make explicit use of specific advanced instructions can often see significant performance benefits. When programming in C and C++, many advanced instructions are available through special functions called intrinsics (see Table~\ref{ref:simdinstructions}). Intrinsics enable programmers to taylor their code to the microarchitecture of their processor without writing assembly code. Intrinsics are supported by most C/C++ compilers on x64 platforms including GNU~GCC, the Intel C++ Compiler, Microsoft Visual Studio and LLVM's compilers. We find it convenient to express our algorithms using intrinsics:
\begin{itemize}
\item  Intrinsics are portable across a wide range of compilers.
\item They  allow us to write all of our code in a single language (C).
\item They leave technical details such as the assignment of registers to the compiler. 
\end{itemize}

Many of the SIMD instructions and intrinsics are straightforward. For example, the \intrinsic{\_mm\_or\_si128} intrinsic (and its corresponding \opcode{por} instruction) takes two 128-bit registers and outputs a new 128-bit register made of the bitwise OR of the inputs.

The first vector instructions on x64 processors used 128-bit vectors (starting with the Pentium~4). Recent commodity x64 processors from Intel (Haswell, Broadwell, Skylake, Kaby Lake) and AMD (Carrizo, Ryzen) have 256-bit vectors with the AVX and AVX2 instructions sets.\footnote{Intel released its first AVX2 processor in 2013 using its Haswell microarchitecture.} However, these vectors should be regarded as pairs of 128-bit vectors (each of them called a ``lane''). Indeed, most AVX/AVX2 instructions cannot move data from one 128-bit lane to another. One exception is the \intrinsic{\_mm256\_permutevar8x32\_epi32} intrinsic which we may use to move or copy 32-bit words from any location in the 32-byte vector to any other.

The  \opcode{pshufb}  instruction shuffles the
input bytes  into a new vector containing the same byte values. Given an input register $v$ and a control mask $m$, as vectors of sixteen~bytes, it outputs a new vector $(v_{m_0},v_{m_1},v_{m_2},v_{m_3}, \ldots, v_{m_{15}})$ (assuming that $0 \leq m_i < 16$ for $i=0,1,\ldots, 15$).\footnote{ARM Neon has similar instruction
\opcode{vtbl}, AltiVec similarly defines \opcode{vperm}.} The AVX2 instruction set contains an upgraded version of this instruction (\opcode{vpshufb} with the intrinsic \intrinsic{\_mm256\_shuffle\_epi8}) that does the same shuffling operation inside each of the two 16-byte lanes of a 32-byte register.

Though we may see the \opcode{pshufb} and \opcode{vpshufb} instructions as  ``shuffling'' instructions (moving the bytes around), another useful interpretation is that of a table lookup. 
That is, with the function call \intrinsic{\_mm256\_shuffle\_epi8}\texttt{(a,b)}, we can effectively treat \texttt{a} as a lookup table, and \texttt{b} as 4-bit indexes. Hence, we can use \opcode{pshufb} and \opcode{vpshufb} to implement maps with the benefit that there are only two registers involved and no other memory access. Moreover, these instructions are inexpensive: they have a throughput of one instruction per cycle on recent Intel processors.

The vectors instructions we use, including \opcode{vpshufb}, have low latency (usually 1~CPU cycle) and a high throughput~\cite{fog2016instruction}. 
Depending on the compilation settings, especially optimization level and the selected CPU target, a compiler might emit different sequences of instructions to express a given sequence of intrinsics.
The intrinsics used to initialize vectors---such as \intrinsic{\_mm256\_set1\_epi16}---can be implemented by the compiler as a load instruction (\opcode{vmovdqa}). However,  within a loop, vector initializations are often optimized away when the compiler recognizes the resulting vector as a constant. The vector might then be initialized once  into a vector register that gets reused. The AVX2 instruction set has access to  sixteen 32-byte (YMM) registers.

\begin{table}
\tbl{\label{ref:simdinstructions}Intel intrinsics and instructions on  x64 processors.}{%
 \centering
\begin{tabular}{lccp{2.6in}}
\toprule
intrinsic & bits & instruction & description  
\\\midrule
\intrinsic{\_mm\_set1\_epi8} & 128 & --- & create a vector containing 16~identical bytes \\ 
\intrinsic{\_mm\_or\_si128} & 128 & \opcode{por} & bitwise OR \\
\intrinsic{\_mm256\_or\_si256} & 256 & \opcode{vpor} & bitwise OR \\
\intrinsic{\_mm\_and\_si128} & 128 & \opcode{pand} & bitwise AND \\
\intrinsic{\_mm256\_and\_si256} & 256 & \opcode{vpand} & bitwise AND \\
\intrinsic{\_mm\_cmpeq\_epi8} & 128 & \opcode{pcmpeqb} & compare 16 pairs of bytes, outputting 0xFF on equality and 0x00 otherwise \\ 
\intrinsic{\_mm256\_cmpeq\_epi8} & 256 & \opcode{vpcmpgtb} & compare 32 pairs of bytes, outputting 0xFF on equality and 0x00 otherwise \\ 
\intrinsic{\_mm\_movemask\_epi8} & 128 & \opcode{pmovmskb} & construct a 16-bit integer from the most significant bits of 16~bytes\\
\intrinsic{\_mm256\_shuffle\_epi8} & 256 & \opcode{vpshufb} & shuffle two lanes of 16 bytes \\
\intrinsic{\_mm\_popcnt\_u64} & 64 & \opcode{popcnt} & return the number of 1s in a 64-bit word (population count)\\
\intrinsic{\_mm\_loadu\_si128} & 128 & \opcode{movdqu} & load 16 bytes from memory into a vector register \\
\intrinsic{\_mm256\_loadu\_si256} & 256 &  \opcode{vmovdqu}& load 32 bytes from memory into a vector register \\
\intrinsic{\_mm256\_set\_epi8} & 256 & -- & load 32 specified bytes into a vector register \\
\intrinsic{\_mm256\_setr\_epi8} & 256 & -- & load 32 specified bytes into a vector register (listed in reverse order) \\
\intrinsic{\_mm256\_set1\_epi8} & 256 & -- & create a vector register with the same 8-bit word repeated 32~times \\
\intrinsic{\_mm256\_set1\_epi32} & 256 & -- & create a vector register with the same 32-bit word repeated four times \\
\intrinsic{\_mm256\_maskload\_epi32} & 256 & \opcode{vpmaskmovd} & load 32 bytes from memory into a vector register, omitting values some values according to the provided mask\\
\intrinsic{\_mm\_storeu\_si128} & 128 &  \opcode{movdqu} & write a 16-byte vector register to memory \\ 
\intrinsic{\_mm256\_storeu\_si256} &  256 &  \opcode{vmovdqu}& write a 32-byte vector register to memory \\ 
\intrinsic{\_mm256\_mulhi\_epu16} & 256 & \opcode{vpmulhuw} & multiply 16-bit integers, keeping the high 16~bits of the result \\
\intrinsic{\_mm256\_mullo\_epi16} & 256 & \opcode{vpmullw} & multiply 16-bit integers, keeping the low 16~bits of the result \\
\intrinsic{\_mm256\_subs\_epu8} & 256 & \opcode{vpsubusb} & subtract 8-bit unsigned integers \\
\intrinsic{\_mm256\_add\_epi8} & 256 & \opcode{vpaddb} & add 8-bit integers\\
\intrinsic{\_mm256\_srli\_epi32} & 256 & \opcode{vpsrld} & shift 32-bit words right by $x$ bits \\
\intrinsic{\_mm256\_testz\_si256} & 256 & \opcode{vptest} & AND the two inputs and return 1 if the result is zero \\
\intrinsic{\_mm256\_subs\_epu8} & 256  & \opcode{vpsubusb} & Subtract 8-bit unsigned integers, setting zero when the result would be negative\\
\intrinsic{\_mm256\_maddubs\_epi16} & 256 & \opcode{vpmaddubsw} & Vertically multiply pairs of 8-bit integers, producing intermediate 16-bit integers. Horizontally add adjacent pairs of intermediate signed 16-bit integers. \\
\intrinsic{\_mm256\_madd\_epi16} & 256 & \opcode{vpmaddwd} & Multiply pairs of 16-bit integers, producing intermediate signed 32-bit integers. Horizontally add adjacent pairs of intermediate 32-bit integers. \\
\intrinsic{\_mm256\_permutevar8x32\_epi32} & 256 & \opcode{vpermd} & Shuffle 32-bit integers in across lanes.
 \\
\bottomrule
\end{tabular}}
\end{table}



\section{Vectorized Base64}

Encoding and decoding base64 data involves mapping blocks of three~bytes to blocks of four~ASCII characters. We usually decode  data by accessing each character, one by one.

For better speed, we may want to decode  data using entire vector registers.
Depending on the largest available vector length, we may be able to decode base64 data in sets of  16 (e.g., SSE, ARM Neon, Power AltiVec), 32 (e.g., AVX/AVX2) or even 64 (e.g., AVX-512) characters at once. Similarly, we may want to encode base64 data in blocks corresponding to vector registers. 

  Recent x64 processors have 32-byte vector registers (AVX2/AVX). Thus, for better speed on this popular platform, we may want to read blocks of 32~characters and transform them into 24~decoded bytes in one step, and \emph{vice versa}.

We can vectorize the  encoding in the following manner: 
\begin{enumerate}
\item Load 24~new input bytes in a 32-byte register.
\item Unpack the 24~bytes into thirty-two~6-bit values.
\item Map each of the thirty-two~6-bit values, into its corresponding ASCII character (see Table~\ref{tab:base64table}).
\item Store the resulting 32~bytes.
\end{enumerate}
The  decoding proceeds in reverse:
\begin{enumerate}
\item Load 32~bytes, treating them as ASCII characters.
\item Map each of the 32~ASCII characters to its corresponding 6-bit value.
\item Pack the thirty-two~6-bit values into  24~bytes (within a 32-byte register).
\item Store the 24~new bytes.
\end{enumerate}

This approach assumes that the original data encoded in base64  is divisible by 24~bytes. However, we can  process the remaining data using one of the scalar algorithms from \S~\ref{sec:base64}.

To our knowledge, the first attempt to vectorize base64 encoding and decoding is due to Klomp~\cite{Klomp2014}. It used  more instructions than the streamlined approach we present, and it lacked a performance evaluation.

\paragraph{ASCII Characters and 6-bit Values}
An important step in coding and decoding base64 data is to map
6-bit values (in $[0,64)$) to ASCII characters. Scalar code (see \S~\ref{sec:base64}) often represents and computes these maps using pre-calculated tables. We could proceed in a similar manner for vectorized code. Given a vector register of 32~bytes, for example, we might have to look up 32~values. For better speed in a vectorized setting, we use the fact that the base64 standard uses ASCII codes spanning five continuous ranges of values corresponding to
the upper-case characters ('\texttt{A}' to '\texttt{Z}'), the lower-case characters ('\texttt{a}' to '\texttt{z}'), the ten digits ('\texttt{0}' to '\texttt{9}') and the characters '\texttt{+}' and '\texttt{/}'. See Table~\ref{tab:shifts}. This observation allows us to replace table lookups with some arithmetic and logical operations, and a few \opcode{vpshufb} instructions. Because vector instructions can operate on large blocks of bytes, arithmetic and logical instructions are amortized and relatively inexpensive.
%
%

\begin{table}
\tbl{Offsets used in translation between 6-bit integers in $[0,64)$ and ASCII values: e.g., to convert integers in $[0,25]$ to ASCII code, we must add 65.\label{tab:shifts}}{%
\begin{tabular}{|c|c|r|}
\toprule
6-bit value     & ASCII range   & offset (6-bit to ASCII)\\
\midrule
 0 \ldots 25    & \texttt{A} \ldots \texttt{Z}    & $65$         \\
26 \ldots 51    & \texttt{a} \ldots \texttt{z}   & $71$        \\
52 \ldots 61    & \texttt{0} \ldots \texttt{9}    & $-4$        \\
62              & \texttt{+}             & $-19$       \\
63              & \texttt{/}             & $-16$         \\
\bottomrule
\end{tabular}}
\end{table}

\subsection{Vectorized  Encoding}

There are two  steps in the vectorized  encoding:
\begin{enumerate}
\item Unpacking the 24~bytes into thirty-two~6-bit values.
\item Translating each of the thirty-two~6-bit values, into its corresponding ASCII character.
\end{enumerate}

\subsubsection{Unpacking Procedure}

During the encoding, we load 24~bytes into a 32-byte AVX2 register. Toward the end of the stream, if there are fewer than 32~bytes left, we can use a masked load (with the \intrinsic{\_mm256\_maskload\_epi32} intrinsic) to avoid exceeding memory bounds.

 Unpacking 6-bit values into separate bytes is done in two steps.  
\begin{itemize}
\item We move each of the eight 3-byte chunks of four 6-bit values into separate 32-bit words.
\item Within each 32-bit word, we move the 6-bit values into separate bytes. 
\end{itemize}


We use the byte-shuffling instruction \opcode{vpshufb} to move 3-byte chunks into separate 32-bit words. Unfortunately the
instruction separates shuffling within halves (\textit{lanes}) of the register which may cause difficulties (see \S~\ref{sec:advancedinstructions}). Indeed, suppose that  we were to load 
the eight 3-byte chunks  in a 32-byte vector as 
\texttt{[|----|----|HHHG|GGFF|FEEE|DDDC|CCBB|BAAA]} where \texttt{A\ldots H} denotes bytes from different chunks and the dashes denote unused bytes. We would then need have two 16-byte lanes containing different numbers of 3-byte chunks, with one 3-byte chunk (\texttt{FFF}) overlapping two lanes:  
\texttt{[|FEEE|DDDC|CCBB|BAAA]} and
\texttt{[|----|----|HHHG|GGFF|]}. To avoid this problem, we load the 24-byte data with an offset 4 bytes, thanks to that a register contains four 3-byte chunks in each lane, as shown on Fig.~\ref{fig:enc_load}; the first 4~bytes as well as the last 4~bytes (out of 32~bytes) are to be discarded. The code with intrinsics is given in Fig.~\ref{fig:enc_pack}.

\begin{figure}
{
\begin{tikzpicture}
\draw [fill=gray] (0.00, 0.00) rectangle (0.40, 0.60);
\draw [fill=gray] (0.40, 0.00) rectangle (0.80, 0.60);
\draw [fill=gray] (0.80, 0.00) rectangle (1.20, 0.60);
\draw [fill=gray] (1.20, 0.00) rectangle (1.60, 0.60);
\draw [thin] (1.60, 0.00) rectangle (2.00, 0.60);
\node at (1.80, 0.30) {\tiny $A_0$};
\draw [thin] (2.00, 0.00) rectangle (2.40, 0.60);
\node at (2.20, 0.30) {\tiny $A_1$};
\draw [thin] (2.40, 0.00) rectangle (2.80, 0.60);
\node at (2.60, 0.30) {\tiny $A_2$};
\draw [thin] (2.80, 0.00) rectangle (3.20, 0.60);
\node at (3.00, 0.30) {\tiny $B_0$};
\draw [thin] (3.20, 0.00) rectangle (3.60, 0.60);
\node at (3.40, 0.30) {\tiny $B_1$};
\draw [thin] (3.60, 0.00) rectangle (4.00, 0.60);
\node at (3.80, 0.30) {\tiny $B_2$};
\draw [thin] (4.00, 0.00) rectangle (4.40, 0.60);
\node at (4.20, 0.30) {\tiny $C_0$};
\draw [thin] (4.40, 0.00) rectangle (4.80, 0.60);
\node at (4.60, 0.30) {\tiny $C_1$};
\draw [thin] (4.80, 0.00) rectangle (5.20, 0.60);
\node at (5.00, 0.30) {\tiny $C_2$};
\draw [thin] (5.20, 0.00) rectangle (5.60, 0.60);
\node at (5.40, 0.30) {\tiny $D_0$};
\draw [thin] (5.60, 0.00) rectangle (6.00, 0.60);
\node at (5.80, 0.30) {\tiny $D_1$};
\draw [thin] (6.00, 0.00) rectangle (6.40, 0.60);
\node at (6.20, 0.30) {\tiny $D_2$};
\draw [thin] (6.40, 0.00) rectangle (6.80, 0.60);
\node at (6.60, 0.30) {\tiny $E_0$};
\draw [thin] (6.80, 0.00) rectangle (7.20, 0.60);
\node at (7.00, 0.30) {\tiny $E_1$};
\draw [thin] (7.20, 0.00) rectangle (7.60, 0.60);
\node at (7.40, 0.30) {\tiny $E_2$};
\draw [thin] (7.60, 0.00) rectangle (8.00, 0.60);
\node at (7.80, 0.30) {\tiny $F_0$};
\draw [thin] (8.00, 0.00) rectangle (8.40, 0.60);
\node at (8.20, 0.30) {\tiny $F_1$};
\draw [thin] (8.40, 0.00) rectangle (8.80, 0.60);
\node at (8.60, 0.30) {\tiny $F_2$};
\draw [thin] (8.80, 0.00) rectangle (9.20, 0.60);
\node at (9.00, 0.30) {\tiny $G_0$};
\draw [thin] (9.20, 0.00) rectangle (9.60, 0.60);
\node at (9.40, 0.30) {\tiny $G_1$};
\draw [thin] (9.60, 0.00) rectangle (10.00, 0.60);
\node at (9.80, 0.30) {\tiny $G_2$};
\draw [thin] (10.00, 0.00) rectangle (10.40, 0.60);
\node at (10.20, 0.30) {\tiny $H_0$};
\draw [thin] (10.40, 0.00) rectangle (10.80, 0.60);
\node at (10.60, 0.30) {\tiny $H_1$};
\draw [thin] (10.80, 0.00) rectangle (11.20, 0.60);
\node at (11.00, 0.30) {\tiny $H_2$};
\draw [fill=gray] (11.20, 0.00) rectangle (11.60, 0.60);
\draw [fill=gray] (11.60, 0.00) rectangle (12.00, 0.60);
\draw [fill=gray] (12.00, 0.00) rectangle (12.40, 0.60);
\draw [fill=gray] (12.40, 0.00) rectangle (12.80, 0.60);
\draw [thin] (0.00, -1.75) rectangle (0.40, -1.15);
\node at (0.20, -1.45) {\tiny $A_1$};
\draw [thin] (0.40, -1.75) rectangle (0.80, -1.15);
\node at (0.60, -1.45) {\tiny $A_0$};
\draw [thin] (0.80, -1.75) rectangle (1.20, -1.15);
\node at (1.00, -1.45) {\tiny $A_2$};
\draw [thin] (1.20, -1.75) rectangle (1.60, -1.15);
\node at (1.40, -1.45) {\tiny $A_1$};
\draw [thin] (1.60, -1.75) rectangle (2.00, -1.15);
\node at (1.80, -1.45) {\tiny $B_1$};
\draw [thin] (2.00, -1.75) rectangle (2.40, -1.15);
\node at (2.20, -1.45) {\tiny $B_0$};
\draw [thin] (2.40, -1.75) rectangle (2.80, -1.15);
\node at (2.60, -1.45) {\tiny $B_2$};
\draw [thin] (2.80, -1.75) rectangle (3.20, -1.15);
\node at (3.00, -1.45) {\tiny $B_1$};
\draw [thin] (3.20, -1.75) rectangle (3.60, -1.15);
\node at (3.40, -1.45) {\tiny $C_1$};
\draw [thin] (3.60, -1.75) rectangle (4.00, -1.15);
\node at (3.80, -1.45) {\tiny $C_0$};
\draw [thin] (4.00, -1.75) rectangle (4.40, -1.15);
\node at (4.20, -1.45) {\tiny $C_2$};
\draw [thin] (4.40, -1.75) rectangle (4.80, -1.15);
\node at (4.60, -1.45) {\tiny $C_1$};
\draw [thin] (4.80, -1.75) rectangle (5.20, -1.15);
\node at (5.00, -1.45) {\tiny $D_1$};
\draw [thin] (5.20, -1.75) rectangle (5.60, -1.15);
\node at (5.40, -1.45) {\tiny $D_0$};
\draw [thin] (5.60, -1.75) rectangle (6.00, -1.15);
\node at (5.80, -1.45) {\tiny $D_2$};
\draw [thin] (6.00, -1.75) rectangle (6.40, -1.15);
\node at (6.20, -1.45) {\tiny $D_1$};
\draw [thin] (6.40, -1.75) rectangle (6.80, -1.15);
\node at (6.60, -1.45) {\tiny $E_1$};
\draw [thin] (6.80, -1.75) rectangle (7.20, -1.15);
\node at (7.00, -1.45) {\tiny $E_0$};
\draw [thin] (7.20, -1.75) rectangle (7.60, -1.15);
\node at (7.40, -1.45) {\tiny $E_2$};
\draw [thin] (7.60, -1.75) rectangle (8.00, -1.15);
\node at (7.80, -1.45) {\tiny $E_1$};
\draw [thin] (8.00, -1.75) rectangle (8.40, -1.15);
\node at (8.20, -1.45) {\tiny $F_1$};
\draw [thin] (8.40, -1.75) rectangle (8.80, -1.15);
\node at (8.60, -1.45) {\tiny $F_0$};
\draw [thin] (8.80, -1.75) rectangle (9.20, -1.15);
\node at (9.00, -1.45) {\tiny $F_2$};
\draw [thin] (9.20, -1.75) rectangle (9.60, -1.15);
\node at (9.40, -1.45) {\tiny $F_1$};
\draw [thin] (9.60, -1.75) rectangle (10.00, -1.15);
\node at (9.80, -1.45) {\tiny $G_1$};
\draw [thin] (10.00, -1.75) rectangle (10.40, -1.15);
\node at (10.20, -1.45) {\tiny $G_0$};
\draw [thin] (10.40, -1.75) rectangle (10.80, -1.15);
\node at (10.60, -1.45) {\tiny $G_2$};
\draw [thin] (10.80, -1.75) rectangle (11.20, -1.15);
\node at (11.00, -1.45) {\tiny $G_1$};
\draw [thin] (11.20, -1.75) rectangle (11.60, -1.15);
\node at (11.40, -1.45) {\tiny $H_1$};
\draw [thin] (11.60, -1.75) rectangle (12.00, -1.15);
\node at (11.80, -1.45) {\tiny $H_0$};
\draw [thin] (12.00, -1.75) rectangle (12.40, -1.15);
\node at (12.20, -1.45) {\tiny $H_2$};
\draw [thin] (12.40, -1.75) rectangle (12.80, -1.15);
\node at (12.60, -1.45) {\tiny $H_1$};
\draw [dotted] (1.60, 0.00) -- (0.00, -1.15);
\draw [dotted] (2.80, 0.00) -- (1.60, -1.15);
\draw [dotted] (4.00, 0.00) -- (3.20, -1.15);
\draw [dotted] (5.20, 0.00) -- (4.80, -1.15);
\draw [dotted] (6.40, 0.00) -- (6.40, -1.15);
\draw [dotted] (7.60, 0.00) -- (8.00, -1.15);
\draw [dotted] (8.80, 0.00) -- (9.60, -1.15);
\draw [dotted] (10.00, 0.00) -- (11.20, -1.15);
\draw [dotted] (11.20, 0.00) -- (12.80, -1.15);
\draw [decorate,decoration={brace,amplitude=10pt}] (0.00, 0.60) -- (6.40, 0.60) node [midway,above,yshift=10pt] {\footnotesize lower 128-bit lane of AVX2 register};
\draw [decorate,decoration={brace,amplitude=10pt}] (6.40, 0.60) -- (12.80, 0.60) node [midway,above,yshift=10pt] {\footnotesize higher 128-bit lane};
\node [fill=white] at (6.40, -0.77) {\footnotesize shuffle eight 3-byte words using \texttt{vpshufb}};
\node at (0.20, 0.00) [below] {\footnotesize {0}};
\node at (6.20, 0.00) [below] {\footnotesize {15}};
\node at (6.60, 0.00) [below] {\footnotesize {16}};
\node at (12.60, 0.00) [below] {\footnotesize {31}};
\end{tikzpicture}
}
\caption{\label{fig:enc_load}Encoding: loading and shuffling 24~input~bytes within a 32-byte register within two 16-byte lanes}
\end{figure}

There are two main steps.
\begin{itemize}
\item As previously stated, we reshuffle each of the 16-byte lanes with the \opcode{vpshufb} so that 3-byte chunks go into separate 32-bit words. But we want the bytes to fall in a specific order. 
See Fig.~\ref{fig:enc_shuffle_bytes}.
This choice enables us to finish the bit shuffling using inexpensive arithmetic and logical operations.
This byte-shuffling code is illustrated in  Fig.~\ref{fig:enc_pack}.

\begin{figure}
{
\begin{tikzpicture}
\draw [thin,fill=gray] (0.00, 0.00) rectangle (0.40, 0.60);
\node at (0.20, 0.30) {$c_1$};
\draw [thin,fill=gray] (0.40, 0.00) rectangle (0.80, 0.60);
\node at (0.60, 0.30) {$c_0$};
\draw [thin] (0.80, 0.00) rectangle (1.20, 0.60);
\node at (1.00, 0.30) {$d_5$};
\draw [thin] (1.20, 0.00) rectangle (1.60, 0.60);
\node at (1.40, 0.30) {$d_4$};
\draw [thin] (1.60, 0.00) rectangle (2.00, 0.60);
\node at (1.80, 0.30) {$d_3$};
\draw [thin] (2.00, 0.00) rectangle (2.40, 0.60);
\node at (2.20, 0.30) {$d_2$};
\draw [thin] (2.40, 0.00) rectangle (2.80, 0.60);
\node at (2.60, 0.30) {$d_1$};
\draw [thin] (2.80, 0.00) rectangle (3.20, 0.60);
\node at (3.00, 0.30) {$d_0$};
\draw [thin,fill=lightgray] (3.20, 0.00) rectangle (3.60, 0.60);
\node at (3.40, 0.30) {$b_3$};
\draw [thin,fill=lightgray] (3.60, 0.00) rectangle (4.00, 0.60);
\node at (3.80, 0.30) {$b_2$};
\draw [thin,fill=lightgray] (4.00, 0.00) rectangle (4.40, 0.60);
\node at (4.20, 0.30) {$b_1$};
\draw [thin,fill=lightgray] (4.40, 0.00) rectangle (4.80, 0.60);
\node at (4.60, 0.30) {$b_0$};
\draw [thin,fill=gray] (4.80, 0.00) rectangle (5.20, 0.60);
\node at (5.00, 0.30) {$c_5$};
\draw [thin,fill=gray] (5.20, 0.00) rectangle (5.60, 0.60);
\node at (5.40, 0.30) {$c_4$};
\draw [thin,fill=gray] (5.60, 0.00) rectangle (6.00, 0.60);
\node at (5.80, 0.30) {$c_3$};
\draw [thin,fill=gray] (6.00, 0.00) rectangle (6.40, 0.60);
\node at (6.20, 0.30) {$c_2$};
\draw [thin] (6.40, 0.00) rectangle (6.80, 0.60);
\node at (6.60, 0.30) {$a_5$};
\draw [thin] (6.80, 0.00) rectangle (7.20, 0.60);
\node at (7.00, 0.30) {$a_4$};
\draw [thin] (7.20, 0.00) rectangle (7.60, 0.60);
\node at (7.40, 0.30) {$a_3$};
\draw [thin] (7.60, 0.00) rectangle (8.00, 0.60);
\node at (7.80, 0.30) {$a_2$};
\draw [thin] (8.00, 0.00) rectangle (8.40, 0.60);
\node at (8.20, 0.30) {$a_1$};
\draw [thin] (8.40, 0.00) rectangle (8.80, 0.60);
\node at (8.60, 0.30) {$a_0$};
\draw [thin,fill=lightgray] (8.80, 0.00) rectangle (9.20, 0.60);
\node at (9.00, 0.30) {$b_5$};
\draw [thin,fill=lightgray] (9.20, 0.00) rectangle (9.60, 0.60);
\node at (9.40, 0.30) {$b_4$};
\draw [decorate,decoration={brace,amplitude=10pt}] (0.00, 0.60) -- (3.20, 0.60) node [midway,above,yshift=10pt] {\footnotesize byte 2};
\draw [decorate,decoration={brace,amplitude=10pt}] (3.20, 0.60) -- (6.40, 0.60) node [midway,above,yshift=10pt] {\footnotesize byte 1};
\draw [decorate,decoration={brace,amplitude=10pt}] (6.40, 0.60) -- (9.60, 0.60) node [midway,above,yshift=10pt] {\footnotesize byte 0};
\draw [thin] (0.00, -2.00) rectangle (0.40, -1.40);
\node at (0.20, -1.70) {$b_3$};
\draw [thin] (0.40, -2.00) rectangle (0.80, -1.40);
\node at (0.60, -1.70) {$b_2$};
\draw [thin] (0.80, -2.00) rectangle (1.20, -1.40);
\node at (1.00, -1.70) {$b_1$};
\draw [thin] (1.20, -2.00) rectangle (1.60, -1.40);
\node at (1.40, -1.70) {$b_0$};
\draw [thin] (1.60, -2.00) rectangle (2.00, -1.40);
\node at (1.80, -1.70) {$c_5$};
\draw [thin] (2.00, -2.00) rectangle (2.40, -1.40);
\node at (2.20, -1.70) {$c_4$};
\draw [thin] (2.40, -2.00) rectangle (2.80, -1.40);
\node at (2.60, -1.70) {$c_3$};
\draw [thin] (2.80, -2.00) rectangle (3.20, -1.40);
\node at (3.00, -1.70) {$c_2$};
\draw [thin] (3.20, -2.00) rectangle (3.60, -1.40);
\node at (3.40, -1.70) {$c_1$};
\draw [thin] (3.60, -2.00) rectangle (4.00, -1.40);
\node at (3.80, -1.70) {$c_0$};
\draw [thin] (4.00, -2.00) rectangle (4.40, -1.40);
\node at (4.20, -1.70) {$d_5$};
\draw [thin] (4.40, -2.00) rectangle (4.80, -1.40);
\node at (4.60, -1.70) {$d_4$};
\draw [thin] (4.80, -2.00) rectangle (5.20, -1.40);
\node at (5.00, -1.70) {$d_3$};
\draw [thin] (5.20, -2.00) rectangle (5.60, -1.40);
\node at (5.40, -1.70) {$d_2$};
\draw [thin] (5.60, -2.00) rectangle (6.00, -1.40);
\node at (5.80, -1.70) {$d_1$};
\draw [thin] (6.00, -2.00) rectangle (6.40, -1.40);
\node at (6.20, -1.70) {$d_0$};
\draw [thin] (6.40, -2.00) rectangle (6.80, -1.40);
\node at (6.60, -1.70) {$a_5$};
\draw [thin] (6.80, -2.00) rectangle (7.20, -1.40);
\node at (7.00, -1.70) {$a_4$};
\draw [thin] (7.20, -2.00) rectangle (7.60, -1.40);
\node at (7.40, -1.70) {$a_3$};
\draw [thin] (7.60, -2.00) rectangle (8.00, -1.40);
\node at (7.80, -1.70) {$a_2$};
\draw [thin] (8.00, -2.00) rectangle (8.40, -1.40);
\node at (8.20, -1.70) {$a_1$};
\draw [thin] (8.40, -2.00) rectangle (8.80, -1.40);
\node at (8.60, -1.70) {$a_0$};
\draw [thin] (8.80, -2.00) rectangle (9.20, -1.40);
\node at (9.00, -1.70) {$b_5$};
\draw [thin] (9.20, -2.00) rectangle (9.60, -1.40);
\node at (9.40, -1.70) {$b_4$};
\draw [thin] (9.60, -2.00) rectangle (10.00, -1.40);
\node at (9.80, -1.70) {$b_3$};
\draw [thin] (10.00, -2.00) rectangle (10.40, -1.40);
\node at (10.20, -1.70) {$b_2$};
\draw [thin] (10.40, -2.00) rectangle (10.80, -1.40);
\node at (10.60, -1.70) {$b_1$};
\draw [thin] (10.80, -2.00) rectangle (11.20, -1.40);
\node at (11.00, -1.70) {$b_0$};
\draw [thin] (11.20, -2.00) rectangle (11.60, -1.40);
\node at (11.40, -1.70) {$c_5$};
\draw [thin] (11.60, -2.00) rectangle (12.00, -1.40);
\node at (11.80, -1.70) {$c_4$};
\draw [thin] (12.00, -2.00) rectangle (12.40, -1.40);
\node at (12.20, -1.70) {$c_3$};
\draw [thin] (12.40, -2.00) rectangle (12.80, -1.40);
\node at (12.60, -1.70) {$c_2$};
\draw [decorate,decoration={brace,amplitude=10pt}] (1.60, -1.40) -- (4.00, -1.40) node [midway,above,yshift=10pt] {\footnotesize field $c$};
\draw [decorate,decoration={brace,amplitude=10pt}] (4.00, -1.40) -- (6.40, -1.40) node [midway,above,yshift=10pt] {\footnotesize field $d$};
\draw [decorate,decoration={brace,amplitude=10pt}] (6.40, -1.40) -- (8.80, -1.40) node [midway,above,yshift=10pt] {\footnotesize field $a$};
\draw [decorate,decoration={brace,amplitude=10pt}] (8.80, -1.40) -- (11.20, -1.40) node [midway,above,yshift=10pt] {\footnotesize field $b$};
\draw [decorate,decoration={brace,mirror,amplitude=10pt}] (0.00, -2.00) -- (3.20, -2.00) node [midway,below,yshift=-10pt] {\footnotesize byte 1};
\draw [decorate,decoration={brace,mirror,amplitude=10pt}] (3.20, -2.00) -- (6.40, -2.00) node [midway,below,yshift=-10pt] {\footnotesize byte 2};
\draw [decorate,decoration={brace,mirror,amplitude=10pt}] (6.40, -2.00) -- (9.60, -2.00) node [midway,below,yshift=-10pt] {\footnotesize byte 0};
\draw [decorate,decoration={brace,mirror,amplitude=10pt}] (9.60, -2.00) -- (12.80, -2.00) node [midway,below,yshift=-10pt] {\footnotesize byte 1};
\node [below, fill=white] at (0.00, 0.00) {24-bit chunk};
\node [above, fill=white] at (0.00, -1.40) {shuffled word};
\end{tikzpicture} 
}
\caption{\label{fig:enc_shuffle_bytes}Encoding: from a packed 24-bit chunk, we generate a shuffled 32-bit word by repeating byte~1. }
\end{figure}

\item  It remains to unpack the four 6-bit values $a,b,c,d$ (see Fig.~\ref{fig:enc_shuffle_bits}).
Treating the four bytes as a 32-bit integer, we can isolate the 6-bit values $c$ and $a$ with a bitwise mask (leaving all other values zero). We then need to shift these bits right by 10 and 6 respectively. Because AVX2 lacks a 16-bit variable shift instruction, we
use the \opcode{vpmulhuw} instruction (\intrinsic{\_mm256\_mulhi\_epu16}) that multiplies pairs of
16-bit numbers while storing the most significant 16 bits of the result.
Likewise, we isolate $b$ and $d$ and shift
them left by 4 and 8 respectively. To achieve this shift, we use the multiplication instruction
\opcode{vpmullo} (\intrinsic{\_mm256\_mullo\_epi16}), which  multiplies pairs of
16-bit numbers while storing the least significant 16 bits of the product. \\
Finally, we merge partial results
using a bitwise OR.
\end{itemize}
\begin{figure}\centering
\begin{tabular}{c} 
\begin{lstlisting}
__m256i enc_reshuffle(__m256i input) {
  __m256i in = _mm256_shuffle_epi8(input, _mm256_set_epi8(
        10, 11,  9, 10,  7,  8,  6,  7, 4,  5,  3,  4, 1,  2,  0,  1,
        14, 15, 13, 14, 11, 12, 10, 11, 8,  9,  7,  8, 5,  6,  4,  5
    ));
  __m256i t0 = _mm256_and_si256(in, _mm256_set1_epi32(0x0fc0fc00));
  __m256i t1 = _mm256_mulhi_epu16(t0, _mm256_set1_epi32(0x04000040));
  __m256i t2 = _mm256_and_si256(in, _mm256_set1_epi32(0x003f03f0));
  __m256i t3 = _mm256_mullo_epi16(t2, _mm256_set1_epi32(0x01000010));
  return _mm256_or_si256(t1, t3);
}
\end{lstlisting}
\end{tabular}
\caption{\label{fig:enc_pack}Encoding: mapping 24~input bytes into thirty-two 6-bit values stored in distinct bytes}
\end{figure}

\begin{figure}
{
\input{encoding-shuffle-bits.tikz.tex} 
}
\caption{\label{fig:enc_shuffle_bits}Encoding: arithmetic and logical operations to unpack four 6-bit values to individual bytes from a shuffled input 32-bit word }
\end{figure}

\subsubsection{ASCII Translation}

Once we have all 6-bit values in separate bytes, we need to convert them into ASCII characters. At a high-level, what we need to do is to take the 6-bit value and add the corresponding ``offset'' value from Table~\ref{tab:shifts}. There are five~distinct offset values. For example, if our 6-bit value lies in the interval $[0,25]$, we want to add 65 to it. Yet we want to avoid a potentially expensive lookup.

Instead of using a table lookup, we use a vectorized approach, see Fig.~\ref{fig:enc_pshufb}. The main ingredient is
the \opcode{vpshufb} instruction  (\intrinsic{\_mm256\_permute\_epi8}), which
does a parallel lookup in a destination register using the lower four bits
of bytes from a source register.  Unlike a conventional table lookup, there is little chance of expensive cache misses since the instruction can take two register values. To make use of the \opcode{vpshufb} instruction, we first have to reduce our 6-bit values to 4-bit values.
It suffices to map the 6-bit values to their reduced value in Table~\ref{tab:input_reduce}. For example, we want all values in $[0,25]$ to be mapped to 13, all values in $[26,51]$ to be mapped to 0, and so forth.

For our purposes, we use a \emph{saturated subtraction} (\intrinsic{\_mm256\_subs\_epu8}) of unsigned values: it yields 0 whenever the result
would be less than zero; it might be expressed as $\max(x - y, 0)$.  The
hardware implementation is as fast as a regular subtraction.

By using saturated subtract with value 51, we reduce input values from ranges $[0, 25]$ and
$[26, 51]$ into a single value $0$. The values in $[52,63]$ are mapped to $[1,12]$. Then a comparison 
identifies input values that are less than 26, and adjusts reduced input by assigning code 13 to
the range $[0, 25]$. Finally, an invocation of \opcode{vpshufb} translates
the reduced input into offset values.

\begin{table}
\tbl{Input reduction for encoding\label{tab:input_reduce}}{%
\begin{tabular}{|c|c|r|}
\toprule
6-bit value     & reduced       & offset         \\
\midrule
 0 \ldots 25    & 13            & $65$          \\
26 \ldots 51    & 0             & $71$           \\
52 \ldots 61    & 1 \ldots 10   & $-4$          \\
62              & 11            & $-19$        \\
63              & 12            & $-16$         \\
\bottomrule
\end{tabular}}
\end{table}

\begin{figure}\centering
\begin{tabular}{c} 
\begin{lstlisting}
__m256i toascii(__m256i input) {
    __m256i result = _mm256_subs_epu8(input, b51);
    __m256i less = _mm256_cmpgt_epi8(b26, input);
    result = _mm256_or_si256(result, _mm256_and_si256(less, b13));
    __m256i offsets = _mm256_setr_epi8(
        65, -4, -4, -4, -4, -4, -4, -4, 
        -4, -4, -4, -19,-16, 71, 0, 0,
        65, -4, -4, -4, -4, -4, -4, -4, 
        -4, -4, -4, -19,-16, 71, 0, 0
    );
    result = _mm256_shuffle_epi8(offsets, result);
    return _mm256_add_epi8(result, input);
}
\end{lstlisting}
\end{tabular}
\caption{\label{fig:enc_pshufb}Encoding: function translating byte values in $[0,64)$ to ASCII characters according to the base64 standard. We use the convention that \texttt{bx} is \intrinsic{\_mm256\_set1\_epi8(x)}.}
\end{figure}

\subsection{Vectorized Decoding}
\label{sec:simd}

Our vectorized decoding algorithm is analogous  to our encoding algorithm.  We decompose  decoding in two steps:
\begin{enumerate}
\item Map each of the 32~ASCII characters to its corresponding 6-bit value.
\item Pack the thirty-two~6-bit values into  24~bytes (within a 32-byte register).
\end{enumerate}

Mapping requires both translation from ASCII characters to 6-bit values and verification that all input characters are valid. Treating the ASCII characters as a sequence of byte values, both steps can be efficiently achieved by analyzing the lower and higher nibbles (4-bit halves) of each ASCII character.\footnote{The \emph{lower nibble} of a byte is made of the least significant 4~bits whereas the \emph{higher nibble} is made of the most significant 4~bits.} Our
approach involves only three lookups in AVX2 vectors---each implemented with a single \opcode{vpshufb}---and a few inexpensive vector instructions: a shift, a comparison, a bitwise AND, two additions and a test. 
To make the input character validation inexpensive, we use
a bitset-based approach.\footnote{A bitset is a standard technique to represent any set made of $n$~possible elements using a word of $n$~bits. If the $i^{\mathrm{th}}$ element is present in the set, we set the $i^{\mathrm{th}}$ bit to 1, otherwise we set it to 0.}

Packing is performed on a vector of 6-bit values passed from the previous step (if no errors were found). This step requires just four instructions. First, we shuffle individual bits within 32-bit words using two instructions, forming 24-bit words. Then we change order of individual bytes within the lanes of an AVX2 register with a call to \opcode{vpshufb}. And we finally pack words within lanes into a continuous 24-byte array, which is written to memory.


\subsubsection{ASCII Translation}
\label{sec:shufb}

Input bytes are 
classified to one of five ranges to select the appropriate
\emph{offset}. As per the base64 specification, we have to report errors, defined as a character outside Table~\ref{tab:base64table}. Invalid input characters yield an offset of zero, which indicates errors.  Once we have a vector of offsets, a single subtraction
of that vector and an input vector yields decoded data.

Decoding ASCII characters to 6-bit values would be easier if we did not
need to check for invalid characters. Indeed, if we shift ASCII byte values right by 4 bits, we get 
\begin{itemize}
\item 2 for characters \texttt{+} and \texttt{/}.
\item 3 for characters in \texttt{0} \ldots \texttt{9};

\item either 4 or 5 for characters in \texttt{A} \ldots \texttt{Z} (4 for \texttt{A} \ldots \texttt{O} and 5 for \texttt{P} \ldots \texttt{Z});
\item either 6 or 7 for characters in \texttt{a} \ldots \texttt{z} (6 for \texttt{a} \ldots \texttt{o} and 7 for \texttt{p} \ldots \texttt{z});
\end{itemize}
Thus we can almost derive the offset  from the most significant 4~bits.
To be able to seek the right offset,  we  have to distinguish between the characters '\texttt{+}' and '\texttt{/}'. For this purpose, we can check whether the ASCII character is equal to~'\texttt{/}' (byte value \texttt{0x2F}).
Table~\ref{tab:dec_mapping} gives the desired offset given the most significant 4~bits and least significant 4~bits of an ASCII value.

\begin{table}%
\tbl{Mapping between lower/higher nibble of input bytes and offsets: each column corresponds to a valid most significant 4-bit value (nibble) for a base64 ASCII value whereas each row corresponds to a least significant 4-bit. The table provides the corresponding negated offset from Table~\ref{tab:shifts}. Most signficant 4-bit values outside the range $[2,7]$ do not correspond to valid base64 ASCII characters. For each column, we provide the matching ASCII characters. \label{tab:dec_mapping}}{%
\begin{tabular}{r|rrrrrrc|r}
\toprule
 & 2 & 3 & 4 & 5 & 6 & 7 & $\{0,1\} \cup [8,15]$ &  \multirow{3}{*}{bitset}\\
ASCII & \texttt{+} and \texttt{/} & \texttt{0} \ldots \texttt{9} & \texttt{A} \ldots \texttt{O} & \texttt{P} \ldots \texttt{Z} & \texttt{a} \ldots \texttt{o} & \texttt{p} \ldots \texttt{z} & none & \\
bitset & \texttt{0x01} & \texttt{0x02} & \texttt{0x04} & \texttt{0x08} & \texttt{0x04} & \texttt{0x08} & \texttt{0x10} & \\
\midrule
 0 &  --  & 4 & --   & -65 & --   & -71 & -- & \texttt{0x15}\\
 1 &  --  & 4 & -65 & -65 & -71 & -71  & --& \texttt{0x11}\\
 2 &  --  & 4 & -65 & -65 & -71 & -71  & --& \texttt{0x11}\\
 3 &  --  & 4 & -65 & -65 & -71 & -71  & --& \texttt{0x11}\\
 4 &  --  & 4 & -65 & -65 & -71 & -71  & --& \texttt{0x11}\\
 5 &  --  & 4 & -65 & -65 & -71 & -71  & --& \texttt{0x11}\\
 6 &  --  & 4 & -65 & -65 & -71 & -71  & --& \texttt{0x11}\\
 7 &  --  & 4 & -65 & -65 & -71 & -71  & --& \texttt{0x11}\\
 8 &  --  & 4 & -65 & -65 & -71 & -71  & --& \texttt{0x11}\\
 9 &  --  & 4 & -65 & -65 & -71 & -71  & --& \texttt{0x11}\\
10 &  --  & -- & -65 & -65 & -71 & -71  & --& \texttt{0x13}\\
11 &  19 & -- & -65 & --   & -71 & --    & --& \texttt{0x1A}\\
12 &  --  & -- & -65 & --   & -71 & --    & --& \texttt{0x1B}\\
13 &  --  & -- & -65 & --   & -71 & --    & --& \texttt{0x1B}\\
14 &  --  & -- & -65 & --   & -71 & --    & --& \texttt{0x1B}\\
15 &  16 & -- & -65 & --   & -71 & --    & --& \texttt{0x1A}\\
\bottomrule
\end{tabular}}
\end{table}%

\begin{algorithm}
\begin{algorithmic}[1]
\REQUIRE  an  ASCII characters $x$, with byte value (code point) $\mathrm{ord}(x)$
\STATE $\mathrm{lut\_lo}\leftarrow \{21, 17, 17, 17, 17, 17, 17, 17, 17, 17, 19, 26, 27, 27, 27, 26\}$ (constant)
\STATE $\mathrm{lut\_hi}\leftarrow \{16, 16, 1, 2, 4, 8, 4, 8, 16, 16, 16, 16, 16, 16, 16, 16\}$ (constant)
\STATE $\mathrm{roll}\leftarrow \{0,   16,  19,   4, -65, -65, -71, -71,  0,   0,   0,   0,   0,   0,   0,   0\}$ (constant)

\STATE $h \leftarrow  \lceil \mathrm{ord}(x) / 16 \rceil$
\STATE $c \leftarrow  -1$ if $x$ is '\texttt{/}' and 0 otherwise

\IF {$(\mathrm{lut\_lo}_{\mathrm{ord}(x) \bmod 16}~\mathrm{AND}~\mathrm{lut\_hi}_{h})$ is non-zero}
\STATE we have an invalid character
\ENDIF
\STATE \textbf{return } $\mathrm{ord}(x)  + \mathrm{roll}_{h+c}$
\end{algorithmic}
\caption{Base64 decoding \label{fig:dec_algorithm} for a single ASCII character. The operation AND represents the bitwise logical AND\@.}
\end{algorithm}

Because we need to ensure that only valid ASCII characters are encountered, our algorithm is slightly longer. We give the simplified version that processes a single character in Algorithm~\ref{fig:dec_algorithm} and an example of C source code in Fig.~\ref{fig:dec_translate_lookup}. The main difficulty in Algorithm~\ref{fig:dec_algorithm}---and in our vectorized C code---is to check that the character is part of the expected base64 characters. To solve this problem efficiently, we observe that the byte value of an allowed base64 character as per Table~\ref{tab:base64table} must be such that:
\begin{itemize}
\item Its high nibble must be 2, 3, 4, 5, 6 or 7.
\item When its high nibble is 2, the low nibble must be 11 or 15.
When its high nibble is 3, the low nibble must be in $[0,9]$. When the high nibble is 4 or 6, its low nibble must be non-zero. When the high nibble is 5 or 7, the low nibble must be in $[0,10]$.
\end{itemize} 
Moreover, these conditions are necessary and sufficient for a base64 character to be valid. See Table~\ref{tab:dec_mapping}.
To check quickly that a character is valid, we use a bitset approach to compare the low and high nibble values. It is implemented in Algorithm~\ref{fig:dec_algorithm} with the arrays \texttt{lut\_lo} and  \texttt{lut\_hi}, and a bitwise AND\@. 
We refer the reader to Appendix~\ref{appendix:longandboring} for a detailed description of the bitset approach and of our vectorized code. 

%
%
%
%

\begin{figure}\centering
\begin{tabular}{c} 
\begin{lstlisting}
bool fromascii(__m256i str, __m256i * out) {
        __m256i hi_nibbles  = _mm256_srli_epi32(str, 4);
        __m256i lo_nibbles  = _mm256_and_si256(str, mask_2F);
        __m256i lo    = _mm256_shuffle_epi8(lut_lo, lo_nibbles);
        __m256i eq_2F = _mm256_cmpeq_epi8(str, mask_2F);
        hi_nibbles = _mm256_and_si256(hi_nibbles, mask_2F);
        __m256i hi    = _mm256_shuffle_epi8(lut_hi, hi_nibbles);
        __m256i roll  = _mm256_shuffle_epi8(lut_roll,
                 _mm256_add_epi8(eq_2F, hi_nibbles));
        if (!_mm256_testz_si256(lo, hi)) {
            return false;
        }
        *out = _mm256_add_epi8(str, roll);
        return true;
}
\end{lstlisting}
\end{tabular}
\caption{\label{fig:dec_translate_lookup}Decoding: from ASCII characters to 6-bit values. The function returns \texttt{false} on error. The vectors \texttt{lut\_lo}, \texttt{lut\_hi}, \texttt{lut\_roll} and \texttt{mask\_2F} are provided in the main text.}
\end{figure}

%
%

\subsubsection{Packing Procedure}

After translation, we have 32 values, that must be saved as a 24-byte array.
We pack the 6-bit values using only four instructions. 
\begin{enumerate}
\item We use \opcode{vpmaddubsw} to pack the data within 16-bit words. 
\item We use  \opcode{vpmaddwd} to pack the data within 32-bit words.
\item  Then we use \opcode{vpshufb} to pack within 128-bit lanes. \item Finally, we pack our 24~bytes within a 256-bit vector using \opcode{vpermd}.
\end{enumerate}


\begin{figure}
{
\input{decoding-pack-bits.tikz.tex} 
}
\caption{\label{fig:dec_pack_bits}Decoding: arithmetic operations to pack four 6-bit values stored in  separate bytes to a continuous (packed) 24-bit subword of each 32-bit word }
\end{figure}

We first review how the packing proceeds within 32-bit words (see Fig.~\ref{fig:dec_pack_bits}).
\begin{itemize}
\item We initially have four 6-bit values stored in each of the four bytes, with zeros elsewhere $[0 0 d_5 d_4 d_3 d_2 d_1 d_0 |
        0 0 c_5 c_4 c_3 c_2 c_1 c_0 |
        0 0 b_5 b_4 b_3 b_2 b_1 b_0 |
        0 0 a_5 a_4 a_3 a_2 a_1 a_0 ]$.
\item We call the   \opcode{vpmaddubsw}
(\intrinsic{\_mm256\_maddubs\_epi16}) instruction providing as inputs our data as well as the vector made of the byte values \texttt{0x01, 0x40, 0x01, 0x40,} \ldots. This instruction multiplies pairs of  8-bit integers from its two input vectors, producing intermediate  16-bit integers, it then adds adjacent pairs of 16-bit integers (the first two, the third and the fourth, \ldots). In our case,
we multiply alternatively by \texttt{0x40} (a shift by 6 bits) and \texttt{0x01} (the identity).
By inspection, the output is
$[  0   0   0   0 c_5 c_4 c_3 c_2 |
        c_1 c_0 d_5 d_4 d_3 d_2 d_1 d_0 |
          0   0   0   0 a_5 a_4 a_3 a_2 |
        a_1 a_0 b_5 b_4 b_3 b_2 b_1 b_0 ]$.
\item We then call the \opcode{vpmaddwd}
(\intrinsic{\_mm256\_madd\_epi16}) instruction. It is similar to the  \opcode{vpmaddubsw} instruction: it multiplies pairs of  16-bit integers from its two input vectors, producing intermediate  32-bit integers, it then adds adjacent pairs of 32-bit integers (the first two, the third and the fourth, \ldots). We call it with the vector made of the 16-bit values \texttt{0x0001, 0x1000,} \ldots so that we alternatively shift by 12 bits or compute the identity. By inspection, the result within each 32-bit word is 
$[0 0 0 0 0 0 0 0|
        a_5 a_4 a_3 a_2 a_1 a_0 b_5 b_4 |
        b_3 b_2 b_1 b_0 c_5 c_4 c_3 c_2 |
        c_1 c_0 d_5 d_4 d_3 d_2 d_1 d_0 ]$.
\end{itemize}
We have effectively packed four 6-bit value to 3 bytes, within each 32-bit word, using only two instructions.

Within each 128-bit lane, we can then shuffle the bytes to pack the data bytes at the beginning of the lanes using \opcode{vpshufb} (\intrinsic{\_mm256\_shuffle\_epi8}). We shall have three 32-bit word filled with data within each lane.  Though normally we cannot move data between lanes, the \opcode{vpermd} (\intrinsic{\_mm256\_permutevar8x32\_epi32}) instruction is a useful exception. It can shuffle 32-bit words across lanes, allowing us to get the final result.  Fig.~\ref{fig:dec_pack_bytes} gives a summary overview of how 
the data is moved within an AVX2 register. 
We provide our C implementation  in Fig.~\ref{fig:dec_packing}.

\begin{figure}
{
\input{decoding-pack-bytes.tikz.tex} 
}
\caption{\label{fig:dec_pack_bytes}Decoding: steps required to convert from 32 6-bit indices into a continuous (packed) array of 24~bytes}
\end{figure}

\begin{figure}\centering
\begin{tabular}{c} 
\begin{lstlisting}
__m256i dec_reshuffle(__m256i in) {
    __m256i merge_ab_and_bc = _mm256_maddubs_epi16(in, _mm256_set1_epi32(0x01400140));
    __m256i out = _mm256_madd_epi16(merge_ab_and_bc, _mm256_set1_epi32(0x00011000));
    out = _mm256_shuffle_epi8(out, _mm256_setr_epi8(
        2, 1, 0, 6, 5, 4, 10, 9, 8, 14, 13, 12, -1, -1, -1, -1,
        2, 1, 0, 6, 5, 4, 10, 9, 8, 14, 13, 12, -1, -1, -1, -1
    ));
    return _mm256_permutevar8x32_epi32(out, _mm256_setr_epi32(0, 1, 2, 4, 5, 6, -1, -1));
}
\end{lstlisting}
\end{tabular}
\caption{\label{fig:dec_packing}Decoding: packing the 6-bit values into full bytes}
\end{figure}

\section{Performance Evaluation}
\label{sec:performance}

We implemented our software in C. All tested code has error handling in the sense that unexpected characters are detected.
We use a Linux server with an Intel i7-6700 processor running at \SI{3.4}{GHz}.
This Skylake processor has \SI{32}{kB} of L1 data cache and \SI{256}{kB} of L2 cache per core with \SI{8}{MB} of L3 cache.
The machine has \SI{32}{GB} of RAM (DDR4~2133, double-channel). There is no disk access during our tests. We disabled Turbo Boost and set the processor
to run at its highest clock speed. We used the processor's time stamp counter (\opcode{rdtsc} instruction~\cite{intelbenchmark})
to estimate the number of cycles. Our software is freely available
 (\url{https://github.com/lemire/fastbase64})
and was  compiled using the GNU GCC~5.3 compiler with  the ``\texttt{-O3 -march=native}'' flags.

In addition to our own implementation, we also use the Linux base64 codec (from the 4.10 release), the base64 codec from the QuickTime Streaming Server\footnote{\url{https://opensource.apple.com/source/QuickTimeStreamingServer/QuickTimeStreamingServer-452/CommonUtilitiesLib/base64.c}}, as well as the base64 codec used within the Google Chrome browser (part of Chromium release 56.0.2924.87)~\cite{Galbreathfast}.
In a cross-browser comparison, N\"agele found that the Chrome browser
has the best base64 performance~\cite{Nagele:Thesis:2015}.
All of the codecs check for invalid characters. All codecs assume
that the input is free from white-space characters, except that 
the Linux codec decoder ignores line feed characters (\verb+\n+) within the base64 encoded stream as long as they occur between two blocks of four~characters.
Klomp produced a library that supports SIMD-accelerated decoding~\cite{Klompgithub2014}. The AVX2 code from his library is similar to ours, but it includes more instructions and it is slower ($\approx 2 \times$). 

To test on realistic data, we included standard images from image processing (Lena, peppers and mandril), a short quote from the novel Moby Dick (``Call me Ishmael (\ldots) the same feelings towards the ocean with me.''), a Google logo found to be base64 encoded in the Google search page and a set of icons found base64 encoded in the Bing search page. To ensure reproducibility, all our test data and benchmarking software is available along with our codecs.

Table~\ref{tab:perf} presents our results in number of cycles per input byte, for single-threaded execution. To make sure our results are reliable, we repeat each test 500~times and check that the minimum and the average cycle counts are within 1\% of each other.
We report the minimum cycle count divided
by the number of bytes in the input. Our AVX2 decoder is nearly an order of magnitude faster than the fastest of the conventional decoder (Chrome).

\begin{table}%
\tbl{Decoding performance in CPU cycles per input Base64 byte. We include the number of cycles per byte required to \texttt{memcpy} the base64-encoded data.\label{tab:perf}}{%
\begin{tabular}{|lr|S|SSSSS|}
\toprule
Source & bytes  & {\texttt{memcpy}} & {Linux} & {QuickTime} & {Chrome} & {AVX2 (Klomp)}  & {AVX2} \\\midrule
lena [jpg] &  \num{141020} &\tablenum{0.09} &  \tablenum{20} &  \tablenum{3.1} & \tablenum{1.8} & \tablenum{0.43} & \tablenum{0.21}\\
peppers [jpg] & \num{12640}&\tablenum{0.03} & \tablenum{15.5} & \tablenum{3.1} & \tablenum{1.8}& \tablenum{0.44} & \tablenum{0.21}\\
mandril [jpg] & \num{329632}&\tablenum{0.11} & \tablenum{20} & \tablenum{3.1} & \tablenum{2.1}& \tablenum{0.43} & \tablenum{0.21}\\
Moby Dick [text] & \num{1484}&\tablenum{0.04} & \tablenum{3.6} &  \tablenum{3.2} &  \tablenum{1.8} & \tablenum{0.53}& \tablenum{0.27}\\
google logo [png] & \num{2357}&\tablenum{0.05} & \tablenum{4.2} &  \tablenum{3.1} & \tablenum{2.1} & \tablenum{0.47}& \tablenum{0.23}\\
bing social icons [png] & \num{1355}&\tablenum{0.04} & \tablenum{3.9} &  \tablenum{3.2} &  \tablenum{2.1} &   \tablenum{0.46} &\tablenum{0.23}\\

\bottomrule
\end{tabular}}

\end{table}%

\begin{figure}\centering
\subfloat[Encoding\label{fig:encoding}]{
\includegraphics[width=0.7\columnwidth]{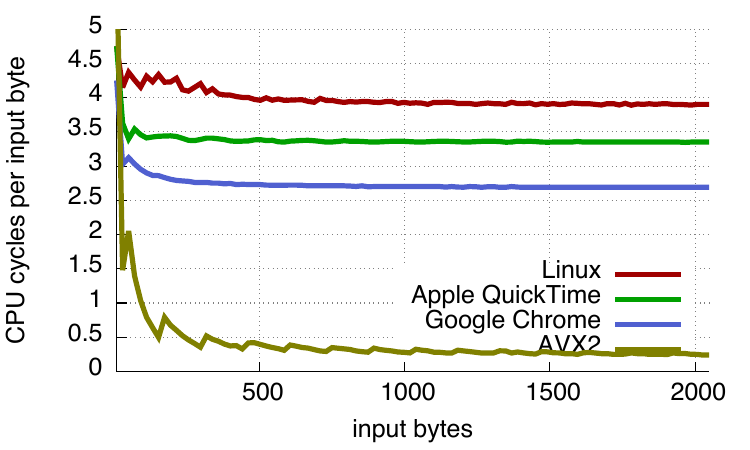}
}\\
\subfloat[Decoding\label{fig:decoding}]{
\includegraphics[width=0.7\columnwidth]{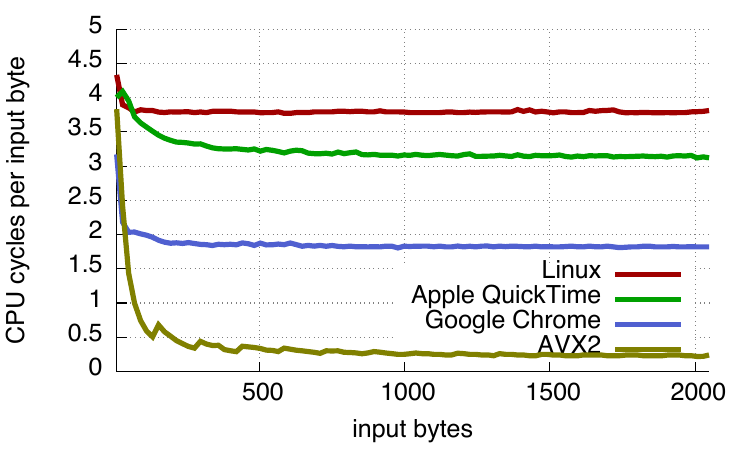}
}
\caption{\label{fig:various}Performance comparison between base64 codecs using random inputs of various lengths}

\end{figure}

To appreciate the performance of the encoding and decoding given inputs having various lengths, we generated random binary data and benchmarked  the time required to encode and decode it (see Fig.~\ref{fig:various}). For sufficiently large inputs, the AVX2 codec uses about a quarter of a cycle per input byte during encoding and decoding. The fastest alternative codec (from Google Chrome) uses about 2.7~cycles per input byte for encoding and about 1.8~cycles per input byte for decoding. Thus the AVX2 codec is eleven times faster at encoding and over seven times faster at decoding than the fastest alternative under consideration in our tests for large inputs. 

Our AVX2 codec falls back on scalar functions when there are fewer than 28~input characters left to encode, or fewer than 45~bytes left to decode. As long as we process at least $\approx 200$~bytes, the AVX2 codec maintains a much higher speed because a sizeable fraction of the data is decoded using SIMD instructions. It is only for short inputs (less than 100~bytes) that the AVX2 codec loses its benefits, achieving a performance no better than scalar functions.

We also analyzed the assembly code produced by our compiler. 
\begin{itemize}
\item 
For encoding 24~input bytes into 32~output bytes (to ASCII), our code uses 11~vector instructions (excluding one load and one store): 
two~\opcode{vpshufb}, two~\opcode{vpand}, one~\opcode{vpor},
 one~\opcode{vpaddb}, one~\opcode{vpmulhuw}, one~\opcode{vpmullw}, one~\opcode{vpsubusb}, one~\opcode{vpsubb} and one~\opcode{vpcmpgtb}. For long inputs, we use about 0.25~cycles per input bytes, or about 6~cycles per 24~input bytes. Thus we execute about 11~vector instructions per 6~cycles, not counting a load and a store, as well as a few scalar instructions.
 \item 
 For decoding 32~input bytes (in ASCII) to 24~output bytes, we use 14~vector instructions (excluding one load and one store): 
four~\opcode{vpshufb},
two~\opcode{vpand},
one~\opcode{vpsrld},
one~\opcode{vpaddb},
one~\opcode{vptest},
one~\opcode{vpcmpeqb},
one~\opcode{vpaddb},
one~\opcode{vpmaddubsw},
one~\opcode{vpmaddwd}, and
one~\opcode{vpermd}. For long inputs, we again use about 0.25~cycles per input bytes, or about 8~cycles per 32~input bytes. Thus execute about 14~vector instructions in 8~cycles, not counting a load and a store and some scalar instructions.
\end{itemize} 
Counting the necessary load and stores, we  need 13~vector instructions to encode 24~input bytes, and we need 16~vector instructions to decode 32~input bytes.
Processors execute complex machine instructions using low-level instructions called \microops{}, each of the vector instructions we considered counts for a single \microop{}~\cite{fog2016instruction}. We require about 0.5~\microops{} per cycle to encode or decode, using our AVX2 codec.
As a comparison, we analyzed the assembly code produced by the fast decoder used by Google Chrome. To decode a single block of four bytes, it requires about 20~\microops{} or about 5~\microops{} per input byte. Thus we can largely explain our good results by the fact that we use fewer instructions per input byte.

\section{Conclusion}

By designing algorithms for vector instructions, we reduced the number of instructions necessary to encode or decode base64 data. Consequently, we showed that commonplace vector instructions (AVX2) can boost  encoding and decoding speeds by roughly an order of magnitude.  The end product spans a few dozens of lines of code and can be readily integrated into existing systems, such as browsers, runtime libraries, and servers.

A vectorized base64 codec can be written for  ARM processors supporting the NEON instructions.
We can preserve the fast algorithms, changing the instructions as needed
(\opcode{vpshufb} becomes \opcode{vtbl} and so forth). Klomp 
reports a $4 \times$ performance gain on an iPhone~\cite{Klompgithub2014} with ARM NEON instructions.
However, there is a  variety of ARM hardware and software configurations. E.g., Srinivasa et al.~\cite{Srinivasa:2017:SBH:3032970.3032982} found 20\% performance differences between otherwise identical devices due to process variation in the manufacture of ARM CPUs. 
We leave a review of  base64 codecs on ARM processors to future work.

Upcoming Intel processors will support some of the AVX-512 instruction sets along with 512-bit vectors.\footnote{\url{http://0x80.pl/articles/avx512-foundation-base64.html}} Even wider registers should allow for ever better performance. However, AVX-512 is a family of instruction sets, with the most powerful sets (e.g., AVX-512BW and AVX-512VL) enabling further optimizations.
ARM processors are also expected to receive support for wider vectors (e.g., 512-bit or 2048-bit vectors) through   the Scalable Vector Extensions~\cite{sve2017}.

Having to design and support functions for several different vectorized instruction sets is cumbersome. We could ask whether we can achieve similar results with higher-level C++ libraries such as  the Generic SIMD Library~\cite{Wang:2014:SPF:2568058.2568059} or Boost.SIMD~\cite{Esterie:2014:BGP:2568058.2568063}. Programming languages like Swift or Rust also have generic SIMD packages that could be put to good use. 
For comparison, we expect that calling our C implementation from  other languages (e.g., Swift and Rust) would be practical and computationally efficient.

For large inputs (e.g., \SI{36}{MB}), Kopp showed that a GPU implementation could encode base64 data roughly twice as fast as a fast CPU implementation~\cite{Kopp2013}. It could be interesting to further explore the engineering of GPU codecs.

%


\appendix
\section*{APPENDIX}
\section{Advanced Error Checking}

\setcounter{section}{1}

We find it interesting that even if all encoded characters are part of our table of base64 characters, Algorithm~\ref{algo:base64decoding} may still decode data that could not possibly have been encoded in base64~\cite{aws2014}. Indeed, we would need to add the following additional checks on the last four ASCII characters ($C_{n-4},C_{n-3},C_{n-2},C_{n-1}$) to ensure there was an original binary input:
\begin{itemize}
\item The stream must end with zero, one or two padding characters ('\texttt{=}').
\item If there are two padding character, then $(A(C_{n-3}) \times 16) \bmod 256 = 0$ where $A$ is as in Algorithm~\ref{algo:base64decoding}. 
\item If there is one padding character, then $(A(C_{n-2}) \times 64) \bmod 256 = 0$.
\end{itemize}
However, these checks are only needed on the last four characters so that we can add them to any existing decoder without measurably impacting the performance. The base64 standard~\cite{rfc4648} does not require these checks.

\section{White Space}

\label{appendix:white}

The base64 specification~\cite{rfc4648} requires decoders to reject encoded data with unexpected characters:
\begin{displayquote}
Implementations MUST reject the encoded data if it contains characters outside the base alphabet when interpreting base-encoded  data unless the specification referring to this document explicitly  states otherwise.
\end{displayquote}
For some applications,  white-space characters may be present and should be removed prior to processing.

We can assume that the input data is encoded in ASCII, UTF-8 or other ASCII supersets (e.g., ISO~8859-1). There are three commonly used white-space characters that we might want to remove:
\begin{itemize}
\item the space character ('\verb+ +', byte value 0x0a or 32 in decimal),
\item the line-feed character ('\verb+\n+', byte value 0x20 or 10 in decimal),
\item the carriage-return character ('\verb+\r+', byte value 0x0D or 13 in decimal).
\end{itemize}

We can write a scalar function that
removes the spaces  (Fig.~\ref{fig:spaceremove}). Our function works by copying the bytes in place while advancing the pointer by one byte only when the character is not a white-space character. On our 
test platform (see \S~\ref{sec:performance}), we can remove
spaces in~place from a small string (1\,kB) containing a small fraction 
of randomly inserted white-space characters (3\%) at a rate of 
1.6~cycles per byte using this function.  We can manually vectorize the removal of white-space characters for better speed (see Fig.~\ref{fig:simdspaceremove}). In effect, it suffices to compare all input characters with each of the possible white-space characters. We get a resulting mask (\texttt{mask16}) that indicates (with a 1-bit) which characters are white space. From this white-space mask, we can lookup a shuffling mask to reorder the input bytes so that white-space characters are omitted. 
We use the \intrinsic{\_mm\_popcnt\_u64} intrinsic (\opcode{popcnt}), which returns the number of 1-bit contained in a 64-bit word. 
The illustrated code sequence removes \verb|_mm_popcnt_u64(mask16)| white space characters from the input vector, so that the first  \verb|16 - _mm_popcnt_u64(mask16)|~bytes
 of the final vector (\texttt{v}) are free from white space. 
 With this faster code, we get a  
processing rate between 0.3 and 0.4~cycles per input byte. We make the source code of our white-space removal software available online (\url{https://github.com/lemire/despacer}).

\begin{figure}\centering
\begin{tabular}{c} 
\begin{lstlisting}
for(i = 0, p = 0; i < N; i++) {
  uint8_t v = A[i];
  A[p] = v;
  p += table[v];
}
\end{lstlisting}
\end{tabular}
\caption{\label{fig:spaceremove}C function to remove selected byte 
values from an array (\texttt{A}) of \texttt{N} bytes in~place: the \texttt{table} array 
is made of 256~Boolean values (0 and 1) where value 0 indicate that the corresponding byte value is to be removed. The integer \texttt{p} represents the new size of the array. }
\end{figure}

\begin{figure}\centering
\begin{tabular}{c} 
\begin{lstlisting}
__m128i spaces = _mm_set1_epi8(' ');
__m128i newline = _mm_set1_epi8('\n');
__m128i carriage = _mm_set1_epi8('\r');

__m128i v = ... // v contains 16 input bytes
__m128i is_spaces = _mm_cmpeq_epi8(v, spaces);
__m128i is_newline = _mm_cmpeq_epi8(v, newline);
__m128i is_carriage = _mm_cmpeq_epi8(v, carriage);
__m128i anywhite = _mm_or_si128(
        _mm_or_si128(xspaces, xnewline), xcarriage);
   
uint64_t mask16 = _mm_movemask_epi8(anywhite);
v = _mm_shuffle_epi8(v, table + mask16));
\end{lstlisting}
\end{tabular}
\caption{\label{fig:simdspaceremove}Vectorized code using Intel intrinsics to remove selected byte 
values from an array (\texttt{A}) of \texttt{N} bytes from a vector of 16~bytes: the \texttt{table} array 
is made of 65536~vectors of 16~bytes. 
}
\end{figure}

Instead of removing three specific characters ('\verb+ +', '\verb+\n+', \verb+\r+'), we could remove all characters with a byte value less than 33 to save a few instructions. However, there is no SIMD instruction on x64 processors providing an unsigned comparison. A sequence of at least two instructions are needed to achieve an unsigned comparison (e.g., a vectorized unsigned maximum \texttt{pmaxub} followed by an equality check).

\section{Complete Description of the Vectorized ASCII Translation}
\label{appendix:longandboring}
Given ASCII characters, we want to map all of them to a 6-bit value as per Table~\ref{tab:base64table}, while reporting unexpected characters as errors.
For this purpose, we describe the vectorized (using AVX2) version of Algorithm~\ref{fig:dec_algorithm}, as represented in C code by Fig.~\ref{fig:dec_translate_lookup}.

\begin{itemize}
\item We shift by 4 bits all of the ASCII byte values, getting values between 2 and 7 (with the \intrinsic{\_mm256\_srli\_epi32} intrinsic), and store the result in \texttt{hi\_nibbles}. 
\item Defining \texttt{mask\_2F} as the vector made of the byte value \texttt{0x2F} (or '\texttt{/}'), we compare the input data to \texttt{mask\_2F} (with the 
\intrinsic{\_mm256\_cmpeq\_epi8} intrinsic). The result is zero when the character value differs from  \texttt{0x2F}, and otherwise, it is \texttt{0xFF} (or -1 as a signed byte integer). We store it in \texttt{eq\_2F}.
\item Later, we can add \texttt{hi\_nibbles} and \texttt{eq\_2F} (with the  \intrinsic{\_mm256\_add\_epi8} intrinsic) and then seek the offset  (with the  \intrinsic{\_mm256\_shuffle\_epi8} intrinsic) from the vector \texttt{lut\_roll} populated with the offset values 
\begin{itemize}
\item     $0,   16,  19,   4, -65, -65, -71, -71,  0,   0,   0,   0,   0,   0,   0,   0$,
\item     $0,   16,  19,   4, -65, -65, -71, -71,  0,   0,   0,   0,   0,   0,   0,   0$.
\end{itemize}
We store the result in \texttt{roll} which contains the decoded offsets.
This is enough to derive the offsets assuming that the ASCII character values were in range.
\item To check that the ASCII character values are in range, we add a few steps.
To be valid, the 4~most significant bits of the value of an ASCII characters should be 2, 3, 4, 5, 6 or 7. Values  in the set $\{0, 1, 8,
9, 10, 11, 12, 13, 14, 15\}$ are always indicative of an invalid character.

Otherwise, given the most significant  4~bits (nibble), only some least significant 4~bits are allowed:
\begin{itemize}
\item When the most significant 4~bits represent the value 2, then we must have either character '\texttt{+}' (byte value \texttt{0x2B}) or character '\texttt{/}' (byte value \texttt{0x2F}) so the least significant 4~bits must represent either 11 or 15 as an unsigned integer.
\item When the most significant 4~bits represent the value 3, then we must have characters  \texttt{0} \ldots \texttt{9}, with a least significant 4~bits ranging in $[0,9]$.
\item When the most significant 4~bits have the value 4 or 6, then least significant 4~bits should be in $[1,15]$. 
\item When the most significant 4~bits have value 5 or 7, then the least significant 4~bits should be in $[0,10]$.
\end{itemize}
See Table~\ref{tab:dec_mapping}.
We can represent these sets efficiently using bitsets.
We start by mapping the most significant 4~bits to 
an 8-bit word with a single bit set: 2 becomes \texttt{0x01}, 3 becomes \texttt{0x02}, 4 and 6 become  \texttt{0x04}, 5 and 7 become \texttt{0x08}, and all other values become  \texttt{0x10}. 
We represent this map with  the vector \texttt{lut\_hi} made of the following 32~bytes (repeating twice the same 16-byte subvector) in a function call such as \intrinsic{\_mm256\_shuffle\_epi8(lut\_hi, hi\_nibbles)}  where \texttt{hi\_nibbles} is a 32-byte vector made of the most significant 4~bits: 
\begin{itemize}
\item     \texttt{0x10}, \texttt{0x10}, \texttt{0x01}, \texttt{0x02}, \texttt{0x04}, \texttt{0x08}, \texttt{0x04}, \texttt{0x08}, \\
            \texttt{0x10}, \texttt{0x10}, \texttt{0x10}, \texttt{0x10}, \texttt{0x10}, \texttt{0x10}, \texttt{0x10}, \texttt{0x10},
\item     \texttt{0x10}, \texttt{0x10}, \texttt{0x01}, \texttt{0x02}, \texttt{0x04}, \texttt{0x08}, \texttt{0x04}, \texttt{0x08}, \\
            \texttt{0x10}, \texttt{0x10}, \texttt{0x10}, \texttt{0x10}, \texttt{0x10}, \texttt{0x10}, \texttt{0x10}, \texttt{0x10}.
\end{itemize}

Similarly, we map the least significant 4~bits to byte values made of the bitwise OR of the byte values corresponding to unallowed most significant 4~bits. 
\begin{itemize}
\item For example, the value zero is mapped to \texttt{0b10101} (or \texttt{0x15}) which is an indication that the most significant 4~bits are allowed to be 3, 5 and 7. That is we disallow 2 (\texttt{0x01}), 4 and 6 (\texttt{0x04}) as well as all values out of the range $[2,7]$ (\texttt{0x10}). And we have that \texttt{0x15} = \texttt{0x10} OR \texttt{0x04} OR \texttt{0x01}.
\item The least significant 4-bit values in $[1,9]$ allow the most significant 4-bit values in $[3,7]$. So we disallow 2 (\texttt{0x01}) and values outside the range  $[2,7]$ (\texttt{0x10}). Thus we map them to \texttt{0x10} OR \texttt{0x01} which is \texttt{0x11}.
\item If the least significant 4-bit value is 10, then we disallow 2 and 3 (as well as values outside $[2,7]$), so that the bitset is \texttt{0x10} OR \texttt{0x02} OR \texttt{0x01} which is \texttt{0x13}.
\item If the least significant 4-bit  value is 11 or 15, then we disallow  3, 5, 7 and values outside $[2,7]$ so that the bitset is \texttt{0x02} OR  \texttt{0x08} OR \texttt{0x10} which is \texttt{0x1A}.
\item If the least significant  4-bit value is 12, 13 or 14, then we disallow  2, 3, 5, 7 and all values outside  $[2,7]$ so that the bitset is  
\texttt{0x10} OR \texttt{0x02} OR \texttt{0x08} OR \texttt{0x01} which is \texttt{0x1B}.
\end{itemize}
We realize this map with  the 32-byte vector \texttt{lut\_lo}  (repeating twice the same 16-byte subvector) and the function call \texttt{\_mm256\_shuffle\_epi8(lut\_lo, lo\_nibbles)} where \texttt{lo\_nibbles} is a 32-byte vector made of the least significant 4~bits: 
\begin{itemize}
\item               \texttt{0x15}, \texttt{0x11}, \texttt{0x11}, \texttt{0x11}, \texttt{0x11}, \texttt{0x11}, \texttt{0x11}, \texttt{0x11},\\
            \texttt{0x11}, \texttt{0x11}, \texttt{0x13}, \texttt{0x1A}, \texttt{0x1B}, \texttt{0x1B}, \texttt{0x1B}, \texttt{0x1A},
\item               \texttt{0x15}, \texttt{0x11}, \texttt{0x11}, \texttt{0x11}, \texttt{0x11}, \texttt{0x11}, \texttt{0x11}, \texttt{0x11},\\
            \texttt{0x11}, \texttt{0x11}, \texttt{0x13}, \texttt{0x1A}, \texttt{0x1B}, \texttt{0x1B}, \texttt{0x1B}, \texttt{0x1A}.
\end{itemize}

To check whether a forbidden character has been detected, we  need to call the intrinsic \intrinsic{\_mm256\_testz\_si256} (\opcode{vptest}) which computes the bitwise AND between two vectors and returns 1 (true) if the value is zero.
\end{itemize}

\begin{example}\label{example:partone}
Let us consider the following 32~input bytes: \texttt{R0lGODlhAQABAIAAAP///wAAACwAAAAA}. The ASCII characters correspond to the following byte values: \\
\texttt{0x52, 0x30, 0x6c, 0x47, 0x4f, 0x44, 0x6c, 0x68,}\\
\texttt{0x41, 0x51, 0x41, 0x42, 0x41, 0x49, 0x41, 0x41,} \\
\texttt{0x41, 0x50, 0x2f, 0x2f, 0x2f, 0x77, 0x41, 0x41,} \\
\texttt{0x41, 0x43, 0x77, 0x41, 0x41, 0x41, 0x41, 0x41}. 
\begin{enumerate}
\item We can map this array of byte values to one made of the most significant 4~bits:\\ 
\texttt{0x5, 0x3, 0x6, 0x4, 0x4, 0x4, 0x6, 0x6,}\\ 
\texttt{0x4, 0x5, 0x4, 0x4, 0x4, 0x4, 0x4, 0x4,}\\
\texttt{0x4, 0x5, 0x2, 0x2, 0x2, 0x7, 0x4, 0x4,}\\
\texttt{0x4, 0x4, 0x7, 0x4, 0x4, 0x4, 0x4, 0x4.}\\
We can compare the original input bytes with \texttt{0x2F} (or '\texttt{/}'), creating a new array that has value \texttt{0xFF} wherever the input character is equal to~'\texttt{/}': \\
\texttt{0x00, 0x00, 0x00, 0x00, 0x00, 0x00, 0x00, 0x00,}\\
\texttt{0x00, 0x00, 0x00, 0x00, 0x00, 0x00, 0x00, 0x00,}\\
\texttt{0x00, 0x00, 0xFF, 0xFF, 0xFF, 0x00, 0x00, 0x00,}\\ 
\texttt{0x00, 0x00, 0x00, 0x00, 0x00, 0x00, 0x00, 0x00.}\\
We can add this value to the most significant 4~bits (modulo 256) to get  \\
\texttt{0x5, 0x3, 0x6, 0x4, 0x4, 0x4, 0x6, 0x6,}\\
\texttt{0x4, 0x5, 0x4, 0x4, 0x4, 0x4, 0x4, 0x4,}\\
\texttt{0x4, 0x5, 0x1, 0x1, 0x1, 0x7, 0x4, 0x4,}\\
\texttt{0x4, 0x4, 0x7, 0x4, 0x4, 0x4, 0x4, 0x4.}\\ 
        We can then seek these 4-bit values (treating them as indexes) in the table  $0,   16,  19,   4, -65, -65, -71, -71,  0,   0,   0,   0,   0,   0,   0,   0$ (e.g., index 0 gets value 0, index 1 gets value 16, \ldots). We get the offsets\\ 
        \verb+-65,   4, -71, -65, -65, -65, -71, -71,+\\ 
        \verb+-65, -65, -65, -65, -65, -65, -65, -65,+\\ 
        \verb+-65, -65,  16,  16,  16, -71, -65, -65,+\\
        \verb+-65, -65, -71, -65, -65, -65, -65, -65.+\\
We can add these offsets to the input byte values to get the decoded 6-bit values: \\
\verb+17, 52, 37,  6, 14,  3, 37, 33,+\\
\verb+ 0, 16,  0,  1,  0,  8,  0,  0,+\\
\verb+ 0, 15, 63, 63, 63, 48,  0,  0,+\\ 
\verb+ 0,  2, 48,  0,  0,  0,  0,  0.+\\
  We can compare the original input string (\texttt{R0lGODlhAQABAIAAAP///wAAACwAAAAA}) with Table~\ref{tab:base64table} and observe that it is, indeed, the expected array. That is, we have successfully mapped the ASCII characters to 6-bit values in $[0,64)$.
\item We also need to check that all of the original ASCII characters are allowed.
This requires two lookups. 
\begin{enumerate}
\item 
First, we take the array of the most significant 4-bit values, and, treating them as 4-bit indexes, we look them up in the table\\ 
\texttt{0x10, 0x10, 0x01, 0x02, 0x04, 0x08, 0x04, 0x08,}  \\
\texttt{0x10, 0x10, 0x10, 0x10, 0x10, 0x10, 0x10, 0x10.} \\
    E.g., index 0 and 1 become \texttt{0x10}, index 2 becomes \texttt{0x01}, and so forth.  From \\
    \texttt{0x5, 0x3, 0x6, 0x4, 0x4, 0x4, 0x6, 0x6,}\\ 
    \texttt{0x4, 0x5, 0x4, 0x4, 0x4, 0x4, 0x4, 0x4,}\\ 
    \texttt{0x4, 0x5, 0x2, 0x2, 0x2, 0x7, 0x4, 0x4,}\\ 
    \texttt{0x4, 0x4, 0x7, 0x4, 0x4, 0x4, 0x4, 0x4,}\\ 
    we get the array \\
    \texttt{0x8, 0x2, 0x4, 0x4, 0x4, 0x4, 0x4, 0x4,}\\ 
    \texttt{0x4, 0x8, 0x4, 0x4, 0x4, 0x4, 0x4, 0x4,}\\ 
    \texttt{0x4, 0x8, 0x1, 0x1, 0x1, 0x8, 0x4, 0x4,}\\ 
    \texttt{0x4, 0x4, 0x8, 0x4, 0x4, 0x4, 0x4, 0x4.}\\ 
    All integers in this array have a single bit set (i.e., they are powers of two).
\item 
From the original input byte values, we can extract an array made of the least significant 4-bits: \\
\texttt{0x2, 0x0, 0xC, 0x7, 0xF, 0x4, 0xC, 0x8,}\\
\texttt{0x1, 0x1, 0x1, 0x2, 0x1, 0x9, 0x1, 0x1,}\\
\texttt{0x1, 0x0, 0xF, 0xF, 0xF, 0x7, 0x1, 0x1,}\\ 
\texttt{0x1, 0x3, 0x7, 0x1, 0x1, 0x1, 0x1, 0x1.}\\
Treating these 4-bit values as indexes, 
we must then look them up in the array \\
\texttt{0x15, 0x11, 0x11, 0x11, 0x11, 0x11, 0x11, 0x11,}\\ 
\texttt{0x11, 0x11, 0x13, 0x1A, 0x1B, 0x1B, 0x1B, 0x1A}\\ 
to get\\
\texttt{0x11, 0x15, 0x1B, 0x11, 0x1A, 0x11, 0x1B, 0x11,}\\
\texttt{0x11, 0x11, 0x11, 0x11, 0x11, 0x11, 0x11, 0x11,}\\ 
\texttt{0x11, 0x15, 0x1A, 0x1A, 0x1A, 0x11, 0x11, 0x11,}\\ 
\texttt{0x11, 0x11, 0x11, 0x11, 0x11, 0x11, 0x11, 0x11.}
\end{enumerate}
 It remains the take the bitwise AND of the two generated array, and we get the zero vector, verifying that all input characters were allowed.
\end{enumerate}
\end{example}

\begin{acks}
The authors would like to thank M.~Howard, A.~Klomp, and N.~Kurz for their software contributions.
\end{acks}

\bibliographystyle{ACM-Reference-Format-Journals}
\bibliography{base64}






\end{document}